\documentclass[twocolumn,showpacs,amsmath,amssymb,showkeys,prd,10pt,nofootinbib,superscriptaddress]{revtex4-1}

\usepackage{epsfig}
\usepackage{latexsym}
\usepackage{graphicx}
\usepackage{amsfonts}
\usepackage{amsmath}
\usepackage{natbib}
\usepackage{xcolor}
\usepackage{wasysym}
\usepackage{bm}
\usepackage{array}

\newcommand{\etal}{{et~al.}}

\newcommand{\biceptwo}{{\sc{bicep2}}}
\newcommand{\keck}{{\sc{Keck}}}
\newcommand{\sptpol}{{\sc SPTpol}}

\newcommand{\pb}{{\sc polarbear}}
\newcommand{\planck}{Planck}

\def\apj{Astrophys. J.}%
\def\aap{Astron. \& Astrophys.}%
\def\prd{Phys.~Rev.~D}%
\def\prl{Phys.~Rev.~Lett.}%
\def\simlt{\lower.5ex\hbox{$\; \buildrel < \over \sim \;$}}
\def\simgt{\lower.5ex\hbox{$\; \buildrel > \over \sim \;$}}

\begin{document}
\title{Detecting the tensor-to-scalar ratio with the pure pseudospectrum reconstruction of $B$-mode}

\author{ A. Fert\'e}
\email{aferte@roe.ac.uk}
\affiliation{%
Institute for Astronomy, University of Edinburgh, Royal Observatory, Blackford Hill, Edinburgh EH9 3HJ}
\affiliation{%
 Universit\'e Paris-Sud 11, Institut d'Astrophysique Spatiale, UMR8617, Orsay, France, F-91405}
\affiliation{%
CNRS, Orsay, France, F-91405}

 \author{J. Peloton}%
 \email{julien.peloton@apc.univ-paris7.fr}
\affiliation{%
AstroParticule et Cosmologie, Universit\'e Paris Diderot, CNRS/IN2P3, CEA/Irfu, Obs. de Paris, Sorbonne Paris Cit\'e, France 
}

\author{J. Grain}
 \email{julien.grain@ias.u-psud.fr}
\affiliation{%
CNRS, Orsay, France, F-91405}
 \affiliation{%
 Universit\'e Paris-Sud 11, Institut d'Astrophysique Spatiale, UMR8617, Orsay, France, F-91405}

 \author{R. Stompor}%
 \email{radek@apc.univ-paris-diderot.fr}
\affiliation{%
AstroParticule et Cosmologie, Universit\'e Paris Diderot, CNRS/IN2P3, CEA/Irfu, Obs. de Paris, Sorbonne Paris Cit\'e, France 
}

\begin{abstract}
$B$-mode of polarized anisotropies of the cosmic microwave background is a unique and nearly direct probe of primordial inflation, which can constrain the amplitude of the primordial gravity waves. 
However, its detection and precise measurement is made difficult by a minute amplitude of the signal, which has to be discerned from many contributions of non-cosmological origin and reliable estimated in the presence of numerous sources of statistical uncertainties. 
Among these latter, the $E$-to-$B$ leakage, arising as a result of partial sky coverage, has been found to play a key and potentially fundamental role in determining the possible statistical significance with which the primordial $B$-mode signal can be detected.
In this work we employ the pure-pseudo formalism devised to minimise the effects of the leakage on the variance of power spectrum estimates and discuss the limits on the tensor-to-scalar ratio, $r$, that could be realistically set by current and forthcoming measurements of the $B$-mode angular power spectrum. 
We compare those with the results obtained using other approaches: na\"\i{ve} mode-counting, minimum-variance quadratic estimators, and re-visit the question of optimizing the sky coverage of small-scale, suborbital experiments in order to maximize the statistical significance of the detection of $r$. 
We show that the optimized sky coverage is largely insensitive to the adopted approach at least for reasonably compact sky patches. 
We find, however, that the mode-counting overestimates the detection significance by a factor $\sim1.17$ as compared to the lossless maximum variance approach and by a factor $\sim1.25$ as compared to the lossy pure pseudo-spectrum estimator. In a second time, we consider more realistic experimental configurations. With a pure pseudospectrum reconstruction of $B$-modes and considering only statistical uncertainties, we find that a detection of $r\sim0.11$, $r\sim0.0051$ and $r\sim0.0026$ at 99\% of confidence level is within the reach of current sub-orbital experiments, future arrays of ground-based telescopes and a satellite mission, respectively. 
This means that an array of telescopes could be sufficient to discriminate between large- and small-field models of inflation, even if the $E$-to-$B$ leakage is consistently included but accounted for in the analysis. 
However, a satellite mission will be required to distinguish between different small-field models depending on the number of e-folds. 
%\begin{color}{red} [In fact the last two statements about small/large field models should be probably mitigated. In fact in depends on the number of e-folds. Maybe we shoud specify that ?]\end{color}
\end{abstract}


\maketitle

%%%%%%%%%%%%%%%%%%%%%%%%%%%%%%%%%%%%%%%%%%%%%%%%%
\section{Introduction} 
%\paragraph*{Introduction--} 
Primordial gravity waves are expected to be produced during cosmic inflation in addition to scalar perturbations. 
If indeed present, they would leave a characteristic footprint on the polarized anisotropies of the cosmic microwave background (CMB), as they are considered to be essentially the sole source of the so-called primordial $B$-mode residing at the super-horizon scales at the time of the last scattering.
A detection of the $B$-mode angular power spectrum at large angular scales would be then treated as a smoking gun of inflation, while a precise measurement of its amplitude would constrain the energy scale of inflation, or, geometrically speaking, the expansion rate of the Universe during inflation \cite{seljak_zaldarriaga_1997,spergel_zaldarriaga_1997}. 
This amplitude is expressed by the tensor-to-scalar ratio, $r$, defined as the relative power of primordial gravity waves with respect to that of the scalar perturbations at some pivot scale $k_0$, chosen here to be equal to $k_0=0.002$ Mpc$^{-1}$. 
Currently, the most stringent upper bound on $r$ using temperature anisotropies has been derived by the \planck\ collaboration : $r<0.11$ at 95\% CL \cite{planck_res_2013}, while a 
recent joint analysis of the \planck\ and \biceptwo\ polarized data set an upper limit $r_{0.05}<0.12$ at 95\% CL \cite{planck_bicep}.
The measurement of the tensor-to-scalar ratio $r$ could allow to discriminate between different inflationary models. In particular,
if this upper bound $r \sim 0.1$ is indeed realized in nature, this would imply a rather high amount of primordial gravity waves thus favoring large-fields inflationary models \cite{lyth}.

At smaller angular scales, the $B$-mode is dominated by the lensing induced signal. 
This signal is generated by the gravitational lensing of the CMB photons due to the large scale structure \cite{zaldarriaga_seljak_1998}. 
The lensing contribution is well-understood from a theoretical point of view and can be uniquely predicted given the primary $E$-modes anisotropies and the lensing deflection field \cite{lewis-challinor-2006}. 
Such predictions have been been recently confirmed by the \sptpol\ \cite{hanson, sptpol-story-2014} and \pb\ experiments~\cite{pbr1,pbr2}, with also constraints on the CMB lensing $B$-mode power spectrum~\cite{pbr3, sptpol-keisler-2015}.  
The lensing $B$-mode signal does not depend on $r$. It therefore acts as a source of an additional 'noise' masking the primordial, $r$-dependent $B$-mode, and making its detection more difficult. 
Striving for a detection of $r$, one has to either try to remove this lensing signal~\cite{delens} or rely solely on the large angular scales. In this latter case two features of the primordial $B$-mode spectrum are of particular interest as they are anticipated to be particularly prominent. 
These are so-called reionization and recombination bumps peaking at $2\leq\ell\lesssim10$ and at $\ell\sim100$, respectively. \\

Measuring $B$-mode is made even more difficult by the fact that measurements as performed by the majority of current experiments, which scan the sky in order to produce its maps, are straightforwardly expressed only in terms of the Stokes parameters, $Q$ and $U$. 
The $E$- and $B$-mode are mathematically related to the Stokes parameters~\cite{zaldarriaga_seljak_1997,kamionkowski_etal_1997,zaldarriaga_2001} and can be therefore recovered from the observational data. 
This however is only simple, if full-sky data were available. In contrast, realistic CMB experiments provide maps of polarized anisotropies, which only cover a reduced fraction of the celestial sphere, ranging from $\sim1\%$ for balloon-borne and ground-based experiments to $\sim70\%$ for satellite missions. 
In the context of pseudospectrum estimation of the angular power spectra on an incomplete sky part of the $E$-mode signal is unavoidably mislabelled as $B$-modes and vice verse. 
Though such leakages can be corrected on average, the leaked signal inevitably contributes to the sampling variance of the other reconstructed spectrum. 
This dramatically increases the uncertainties of the estimated $B$-mode spectrum since the cosmological $E$-mode is expected to be at least two orders of magnitude higher than the $B$-mode in terms of their power spectrum~\cite{bunn_2002}. 
The nature of the leakages and approaches to their removal were investigated in Ref.~\cite{bunn_etal_2003} and a relevant pseudospectrum estimator, referred to as the pure pseudospectrum estimator, was proposed subsequently in 
Ref.~\cite{smith_2006}. 
This estimator has been thoroughly investigated and extended to include an optimization of the sky apodization~\cite{smith_zaldarriaga_2007}, cross-spectrum approaches~\cite{grain_etal_2009}, and, $TB$ and $EB$ cross-correlations~\cite{grain_etal_2012}. 
Alternative constructions of pseudospectrum estimators correcting for $E$-to-$B$ leakages have been also proposed~\cite{zhao_baskaran_2010,kim_naselsky_2010,kim_2011,bowyer_etal_2011}. 
Nevertheless,  the pure pseudospectrum method has been found the most mature and efficient one, particularly due to its ability of optimizing the sky apodizations~\cite{ferte_etal_2013}, making it a method of choice for many practical applications.
It is worth pointing out that the leakages are indeed ubiquities and correcting for them is as mandatory for small-scale experiments, covering $\sim1\%$ of the sky, as for satellite-like missions, with access to as much as $\sim70\%$ of the sky~\cite{ferte_etal_2013}. 

Though the impact of the $E$-to-$B$ leakage on the variance of the $B$-mode power spectrum is generally acknowledged, it is rarely included in projecting 
performance of  planned CMB experiments or instrumental concepts from the point of view of their setting constraints on the tensor-to-scalar ratio, $r$. Instead,
the major body of work (see~\cite{caligiuri_2014, wu_errard_2014} for some recent examples) in this area is based on simplified mode-counting arguments
(see, however e.g.,~\cite{smith_zaldarriaga_2007, stivoli_etal_2010} for some exceptions). This stemmed mostly from the practical reasons, as the impact of the leakage
is neither calculable analytically nor analysis method independent. \\
%\medskip \\

The objective of this work
is to fill  this gap and present a more systematic study of the impact of the presence of the leakage on the performance forecasts 
of CMB B-mode experiments. The paper consists of two parts. In the first part, Sec. \ref{sec:cap}, we consider idealized observations of azimuthally symmetric sky areas 
with homogenous noise and study differences between performance forecasts derived applying three different approaches for different assumed sky area sizes. 
Subsequently, from these
three different perspectives we revisit the issue of the optimal sky area, which would permit setting the most stringent constraints on the scalar-to-tensor
ratio, $r$, given a fixed length and sensitivity of the experiment. 

In the second part, Sec. \ref{sec:realistic}, we complete those considerations by discussing more realistic sky areas defined for  
three types of experiments: small-scale observations covering $\sim 1$\% of the sky, an array of
ground-based telescope covering $\sim 36$\% and a satellite-like mission capable of delivering up to $71$\% of the foreground clean sky. 
Our conclusions are drawn out in Sec. \ref{sec:conclu}, where we also briefly sketch the implications for constraining inflationary models.

Throughout this work we neglect complications such as polarized diffuse foregrounds, e.g.,~\cite{stivoli_etal_2010, errard_2012} and account for resolved points
sources only by appropriately tailoring the adopted mask. We also assume that no subtraction 
of the lensing $B$-mode has been attempted~\cite{delens}.

%%%%%%%%%%%%%%%%%%%%%%%%%%%%%%%%%%%%%%%%%%%%%%%%%
\section{Measuring the tensor-to-scalar ratio for idealized small-scale experiment}
\label{sec:cap}
\subsection{Experimental setup}
We consider first the case of small-scale experiments in an idealized way. 
The observed part of the celestial sphere is assumed to be azimuthally symmetric, given by a spherical cap. 
We however let vary the sky coverage from 0.5\% to 10\%. 
The noise is an homogeneous, white noise, and its level is fixed at $n_P(1\%)=5.75\mu$K-arcminute for $f_\mathrm{sky}=1\%$ (a typical level for ongoing small-scale experiments). 
For a fixed sensitivity and a fixed time of observation, the noise level (in $\mu$K-arcminute) scales as:
\begin{equation}
n_P(f_\mathrm{sky})=\sqrt{\frac{f_\mathrm{sky}[\%]}{1\%}}\times n_P(1\%).
\end{equation}
The instrumental noise reprojected on the sky thus varies from $4.1\mu$K-arcminute to $18\mu$K-arcminute for an observed fraction of the sky of 0.5\% and 10\%, respectively. Finally, the angular resolution is given by an azimuthally symmetric, gaussian beam with a width of 8 arminutes.

We subsequently investigate the signal-to-noise ratio, $\left(\mathrm{S/N}\right)_r=r/\sigma_r$, as a function of the sky coverage. This will be done considering four values of the tensor-to-scalar ratio: $r=0.07,~0.1,~0.15$ and $0.2$ . 
We note that the last two values are disfavored by the current data, nevertheless we include them in our considerations as they are useful
in demonstrating some of the effects we describe hereafter.
% Despite the current upper bound on the tensor-to-scalar ratio $r<0.11$, we first want to address the efficiency of the methods given a broad range of $r$.
%\begin{color}{red} [How to justify the choice $r=0.2$ ? Theoretical consideration here on the techniques and we want to capture different ranges of $r$ ? The constraint from planck essentially comes from large scales, $\ell<20$, and still room the allow for an "effective" value of $r=0.2$ at the recombination scales ? For r=0.2, there would be still the debate about (nt,r) which could explain such a high value. However, we do not probe the very first multipoles.]\end{color}

\subsection{Fisher matrix formalism}
Translating the uncertainties on the $B$-mode angular power spectrum reconstruction into error bars on the measured tensor-to-scalar ratio, $\sigma_r$, can be done using a Fisher matrix formalism. 

For the rather small observed fractions of the celestial sphere here-considered, the $B$-mode angular power spectrum is reconstructed within bandpowers, labelled $b$ hereafter, with bandwidths $\Delta_b$. The binned power spectrum is given by $\mathcal{C}^{B}_b=\sum_\ell P_{b\ell}~ C^{B}_\ell$, where the binning operator is defined as:
\begin{equation}
	P_{b\ell}=\left\{\begin{array}{ll}
		\displaystyle\frac{\ell(\ell+1)}{2\pi\Delta_b}&~~~\mathrm{if}~\ell\in b, \\
		0&~~~\mathrm{if}~\ell\notin b.
	\end{array}\right.
\end{equation}
(Our specific choice of the binning will be given in Sec. \ref{ssec:powspec}.) The error bars on $r$ are then derived from the Fisher matrix via:
\begin{equation}
(\sigma_r)^{-2}=F_{rr}=\displaystyle\sum_{b,b'}\left(\frac{\partial \mathcal{C}^{B}_b}{\partial r}\right)\left(\mathbf{\Sigma}^{-1}\right)_{bb'}\left(\frac{\partial \mathcal{C}^{B}_{b'}}{\partial r}\right),
\label{eq:fisher}
\end{equation}
with $\mathbf\Sigma_{bb'}=\mathrm{Cov}\left(\widehat{\mathcal{C}}^{B}_b,\widehat{\mathcal{C}}^{B}_{b'}\right)$, which stands for the covariance matrix of the reconstructed, binned angular power spectrum of the $B$-mode. (Note that $\widehat{C}^B_\ell$ denotes the {\it estimator} of the angular power spectrum, $C^B_\ell$.)

The $B$-mode angular power spectrum as a function of $r$ is modeled as:
\begin{equation}
	C^{B}_\ell\left(r\right)=r\times \mathcal{T}^{B}_\ell+\mathcal{T}^{E\to B}_{\ell,\mathrm{lens}},
\end{equation}
with $\mathcal{T}^{B}_\ell$ and $\mathcal{T}^{E\to B}_{\ell,\mathrm{lens}}$ two fiducial angular power spectra, which do not depend on $r$. The former is just obtained as the contribution of primordial gravity waves for $r=1$ (taking into account that the primordial $B$-mode is itself lensed). The latter corresponds to the contribution of primary $E$-mode transferred into $B$-mode because of the gravitational lensing of large scale structures. We do not consider here a potential {\it delensing} of the $B$-mode anisotropies, and such a contribution will be assumed to act as an additional gaussian noise for the measurement of $r$. This is a simplifying assumption since the lensing-induced $B$-mode is non-gaussian leading to an additional, non-gaussian contribution to the covariance \cite{smith_etal_2004}. Gaussianity remains however a good approximation for bandpowers which are narrow enough ($\Delta_b\lesssim100$) \cite{smith_etal_2004}, which is the case in our study.

The covariance matrix $\mathbf\Sigma_{bb'}$ is estimated using three different approaches, as described here.

\subsubsection{Mode-counting}
First, we rely on a na\"\i{ve} mode-counting expression (or so-called $f_\mathrm{sky}$-formula). In this case, the covariance on $\widehat{C}^{B}_\ell$ is approximated by:
\begin{equation}
\mathrm{Cov}\left(\widehat{C}^{B}_\ell,\widehat{C}^{B}_{\ell'}\right)=\frac{2\delta_{\ell,\ell'}}{(2\ell+1)f_\mathrm{sky}}\left(C^{B}_\ell+\frac{N_\ell(f_\mathrm{sky})}{B^2_\ell}\right)^2, \label{eq:mcount}
\end{equation}
with $N_\ell$ the noise power spectrum, $B_\ell$ the beam of the telescope, and, $f_\mathrm{sky}$ the portion of the celestial sphere, which is observed (or kept in the analysis). The noise power spectrum scales linearly with the sky coverage. 

The covariance matrix for the binned power spectrum is thus given by:
\begin{equation}
	\mathbf\Sigma_{bb'}=\left[\displaystyle\sum_{\ell\in b}(P_{b\ell})^2\times\mathrm{Cov}\left(\widehat{C}^{B}_\ell,\widehat{C}^{B}_{\ell}\right)\right]\delta_{b,b'}. 
	\label{eq:mcountcov}
\end{equation}
This is essentially used as a benchmark as such an evaluation of the statistical error bars on the $B$-mode reconstruction underestimates the error bars coming from any numerical methods to be applied to the data.

\subsubsection{Minimum variance quadratic estimator}
Second we consider the error bars that could be incurred by using a minimum variance quadratic estimator \cite{tegmark,tegmark_pol}. The estimator is defined as follows:
\begin{equation}
	\widehat{C}^{B}_\ell=\frac{1}{2}\displaystyle\sum_{\ell'}\mathbf{F}^{-1}_{~~\ell\ell'}\left\{\mathrm{Tr}\left[\mathbf{d}^\dag\left(\mathbf{C}^{-1}\frac{\partial\mathbf{C}}{\partial C^{B}_{\ell'}}\mathbf{C}^{-1}\right)\mathbf{d}\right]-\tilde{N}_{\ell'}\right\}.
	\label{eq:copt}
\end{equation}
In the above, $\mathbf{C}=\left<\mathbf{d}\mathbf{d}^\dag\right>$ is the covariance matrix of the maps of the Stokes parameter, and $\mathbf{d}$ is the column vector composed of $(I,~Q,~U)$ (the trace operation is across pixels). The quantity $\tilde{N}_{\ell'}$ stands for the noise debias. Finally, $\mathbf{F}$ is the Fisher information matrix given by:
\begin{equation}
	\mathbf{F}_{\ell\ell'}=\frac{1}{2}\mathrm{Tr}\left[\frac{\partial\mathbf{C}}{\partial C^{B}_\ell}\mathbf{C}^{-1}\frac{\partial\mathbf{C}}{\partial C^{B}_{\ell'}}\mathbf{C}^{-1}\right].
	\label{eq:fishopt}
\end{equation}
It is then shown that the covariance of the above estimator is given by the inverse of the Fisher matrix, i.e. $\mathrm{Cov}\left(\widehat{C}^{B}_\ell,\widehat{C}^{B}_{\ell'}\right)=\mathbf{F}^{-1}_{~~\ell\ell'}$. We remind that this estimator is precisely built to be the quadratic estimator with the lowest variance.

If the $B$-mode power spectrum is indeed estimated for each multipole, $\ell$ (that is chosing $\Delta_b=1$), this directly gives the following expression for the error bars expected on $r$:
\begin{equation}
(\sigma_r)^{-2}=F_{rr}=\frac{1}{2}\mathrm{Tr}\left[\frac{\partial\tilde{\mathbf{C}}}{\partial r}\tilde{\mathbf{C}}^{-1}\frac{\partial\tilde{\mathbf{C}}}{\partial r}\tilde{\mathbf{C}}^{-1}\right],
\end{equation}
with $\tilde{\mathbf{C}}$ the same covariance matrix {\it but} assuming that only $C^{B}_\ell$ does depend on $r$, in line with our approach consisting in constraining the tensor-to-scalar ratio from the $B$-mode's measurements only\footnote{We note that from the complete definition of $\sigma_r$ given by:
$$
(\sigma_r)^{-2}=\frac{1}{2}\mathrm{Tr}\left[\frac{\partial{\mathbf{C}}}{\partial r}{\mathbf{C}}^{-1}\frac{\partial{\mathbf{C}}}{\partial r}{\mathbf{C}}^{-1}\right],
$$
with $\mathbf{C}$ the covariance matrix assuming that all the angular power spectra do depend on $r$, the equation (\ref{eq:fisher}) is therefore replaced by:
$$
(\sigma_r)^{-2}=\displaystyle\sum_{A,A'}\sum_{\ell,\ell'}\left(\frac{\partial C^{A}_\ell}{\partial r}\right)\left(\mathbf{\Sigma}^{-1}\right)^{AA'}_{\ell\ell'}\left(\frac{\partial C^{A'}_{\ell'}}{\partial r}\right).
$$
In the above, the indices $A,~A'$ runs over $TT,~EE,~BB,~TE,~TB$ and $EB$. The equation (\ref{eq:fisher}) is finally obtained assuming that only $C^{B}_\ell$ in $\mathbf{C}$ does depend on the tensor-to-scalar ratio.}. In the case of azimuthally symmetric patches, the numerical computation of such Fisher matrices (either $\mathbf{F}_{\ell\ell'}$ or $F_{rr}$), can be performed in a reasonable time using the expression found in the appendix F of Ref. \cite{smith_2006}, and by using the {\sc s$^2$hat} package to perform spherical harmonic transforms \cite{s2hat,pures2hat,hupca_etal_2010,szydlarski_etal_2011}. (The use of this massively parallel package allows for a rapid computation of the covariance matrix for large sky coverages.). In the standard case, for brute force calculation the Fisher matrix requires $\mathcal{O}(N_\mathrm{pix}^3)$ operations to be evaluated, but in this calculation, evaluating the spin harmonics by recursion in $\ell$ makes the computational cost as $\mathcal{O}(N_\theta^3 m_\mathrm{max})$, where $N_\theta$ is the number of rings actually used.

Practically speaking, one should nonetheless include the impact of binning, done as follows. First one defines the so-called {\it optimal pseudospectrum}:
\begin{equation}
	\tilde{C}^{\mathrm{(opt)}}_\ell=\mathrm{Tr}\left[\mathbf{d}^\dag\left(\mathbf{C}^{-1}\frac{\partial\mathbf{C}}{\partial C^{B}_{\ell}}\mathbf{C}^{-1}\right)\mathbf{d}\right]-\tilde{N}_{\ell}.
\end{equation}
One easily checks that $\left<\tilde{C}^{\mathrm{(opt)}}_\ell\right>=2\sum_{\ell'}\mathbf{F}_{\ell\ell'}B^2_{\ell'}C^{B}_{\ell'}$ (where we also include the impact of an azimuthally symmetric beam). One then introduces the matrix:
\begin{equation}
	\tilde{\mathbf{F}}_{bb'}=\displaystyle\sum_{\ell\in b}\sum_{\ell'\in b'}P_{b\ell}F_{\ell\ell'}B^2_{\ell'}Q_{b'\ell'},
\end{equation}
with the interpolation operator, $Q_{b\ell}$:
\begin{equation}
	Q_{b\ell}=\left\{\begin{array}{ll}
		\displaystyle\frac{2\pi}{\ell(\ell+1)}&~~~\mathrm{if}~\ell\in b, \\
		0&~~~\mathrm{if}~\ell\notin b.
	\end{array}\right.
\end{equation}
The binned estimator, $\widehat{\mathcal{C}}^{B}_b$, is finally defined as:
\begin{equation}
	\widehat{\mathcal{C}}^{B}_b=\displaystyle\sum_{b'}\tilde{\mathbf{F}}^{-1}_{~~bb'}~\tilde{\mathcal{C}}^{\mathrm{(opt)}}_{b'}
\end{equation}
with $\tilde{\mathcal{C}}^{\mathrm{(opt)}}_{b'}=\sum_{\ell'}P_{b'\ell'}~\tilde{C}^{\mathrm{(opt)}}_{\ell'}$ the binned, optimal pseudospectrum. From that last definition, and making use of Eqs. (\ref{eq:copt}) and (\ref{eq:fishopt}), it is straightforward to show that:
\begin{equation}
	\mathbf{\Sigma}_{bb'}=[\tilde{\mathbf{F}}^{-1}]_{bb_1}\left[P_{b_1\ell_1}~\mathbf{F}_{\ell_1\ell'_1}~P_{b'_1\ell'_1}\right][(\tilde{\mathbf{F}}^{-1})^\dag]_{b'_1b'}, 
	\label{eq:qmlcov}
\end{equation}
where summations over repeated indices (i.e. $b_1,b'_1$ and $\ell_1,\ell'_1$) is implicitly assumed, and $\dag$ means the transpose operation. \\

We note that this way of estimating the uncertainties on the power spectrum reconstruction is also relevant for maximum-likelihood approaches, see e.g. Ref. \cite{bond_etal_1998}.
%, and despite the fact that we do not have an exact azimuthally symmetric patches, this method allows to probe the impact of the off-diagonal terms of the covariance matrix in the estimation of the uncertainty.

\subsubsection{Pure pseudospectrum}
Third, we make use of the {\sc x$^2$pure} code and Monte-Carlo simulations to estimate the covariance matrix expected for the pure pseudospectrum approach. Details on the pure pseudospectrum estimator can be found in Refs. \cite{smith_2006,grain_etal_2009}. In practice, the power spectrum is estimated within bandpower and the covariance matrix reconstructed from the MC simulations is directly $\mathbf{\Sigma}_{bb'}$. The numerical cost of this scales as $\mathcal{O}(N^{3/2}_\mathrm{pix})$, allowing for rapid MC simulations.

For each sky coverage and for each value of $r$ here-considered, we compute optimized sky apodizations to apply to the maps of $Q$ and $U$. Those optimized sky apodizations are described in Refs. \cite{smith_zaldarriaga_2007,grain_etal_2009} and they allow for having the smallest error bars on the $B$-mode power spectrum reconstruction within the context of pure pseudospectrum techniques. Those sky apodizations are a set of spin-0, spin-1 and spin-2 window functions to be applied to the maps of the Stokes parameter. They can be interpreted as the window functions, which make the pure pseudospectrum estimator as close as possible to the minimum-variance quadratic estimator \cite{smith_zaldarriaga_2007}. 

Numerically speaking, computing those optimized sky apodizations may be long, especially for intricate shape of the observed region and/or low level of noise. Using an iterative method, the numerical cost is $\mathcal{O}(N_\mathrm{iter}N^2_\mathrm{pix})$, with $N_\mathrm{iter}$ a number of iterations ranging from few tens to few hundreds for simple patch geometry and noise level considered here (see Sec. III in \cite{grain_etal_2009}). We stress that for a given sky patch, those sky apodizations are to be optimized for each value of the tensor-to-scalar ratio and bin-by-bin. Taking into account the number of bins (see Sec. \ref{ssec:powspec}), the number of sky fractions and the number of values of $r$, which are sampled in this study, this means that 1664 of such sky apodizations have to be computed. Fortunately, in the case of homogeneous noise and patches with relatively simple contours (which is obviously the case for a spherical cap), it was demonstrated in Ref. \cite{grain_etal_2009} that an approximated but numerically fast computation of those sky apodizations in the harmonic domain is possible, and indeed leads to error bars equal to those obtained thanks to a direct, pixel-based computation of the optimized sky apodizations. The numerical cost of this technique is reduced to $\mathcal{O}(N^{3/2}_\mathrm{pix})$ which allows us to derive optimized sky apodizations for each value of the sky coverage, and for each value of the tensor-to-scalar ratio.

\begin{figure*}
\begin{center}
	\includegraphics[scale=0.35]{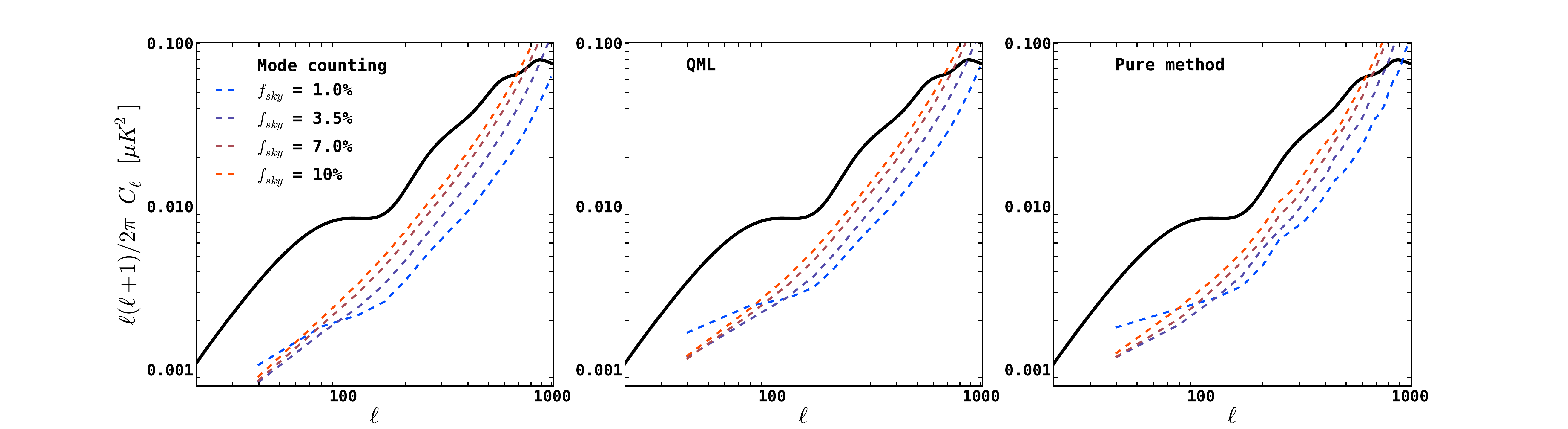}
	\caption{Uncertainties on the reconstructed $B$-mode power spectrum (dashed color curves) for four values of $f_\mathrm{sky}$: 1\%, 3.5\%, 7\% and 10\%. The solid black curve is the input $C^{B}_\ell$ for $r=0.1$. Each panel corresponds to a different approach to derive the covariance matrix $\mathbf{\Sigma}_{bb'}$: mode-counting, minimum variance quadratic estimator, and, pure pseudospectrum estimator (from left to right).}
	\label{fig:powunc1}
\end{center}
\end{figure*}

\subsection{Power spectrum uncertainties}
\label{ssec:powspec}

Our chosen bandpowers for reconstructing $C^{B}_\ell$ are the following. 

The first bin starts at $\ell=20$ and we use a constant bandwidth,  $\Delta_b=40$. 
Our last bin extends up to $\ell=1020$.
The value of the maximum multipole is chosen in order to include all the relevant contributions from the primordial $B$-mode, that is until the lensing $B$-mode be the dominant contribution to the total $B$-mode power.
In addition, for the experimental cases under consideration in this paper, we use beamwidths up to 8 arcminutes, corresponding to a cut-off of $\ell \sim 1300$.
The choice of the bandwidth is mainly motivated by the use of the pure pseudospectrum estimator.
Especially, it is mandatory for the numerical inversion of the mode-mixing matrices to be possible. 
We also note that the bandwidth is wide enough so that the correlations between different bins are nearly uncorrelated in the covariance of pure pseudospectrum estimator.
We discuss the role of the bin width later on.
We stress that the multipoles ranging from $\ell=2$ to $\ell=20$ (corresponding to the reionization peak, and gathered in one bandpower) are actually used in the pure pseudo-$C_\ell$ estimation of $C^{B}_\ell$. 
However, given the limited sky coverages considered here such low multipoles are difficult to estimate and hardly constrained by the data. 
This bin is therefore not included in our analysis of  the signal-to-noise ratio.

The uncertainties on the estimated power spectrum of the $B$-mode as functions of the sky fraction, are shown in Fig. \ref{fig:powunc1} where four selected values of the sky coverage are depicted: 1\%, 3.5\%, 7\% and 10\%. 
The tensor-to-scalar ratio chosen for this figure is $r=0.1$. 
Each panel corresponds to a different approach to derive the covariance matrix, $\mathbf{\Sigma}_{bb'}$: mode-counting, minimum variance quadratic estimator, and, pure pseudospectrum estimator (from left to right).

As expected, we do observe that the lowest error bars are the ones from the mode-counting estimation of the uncertainties, while the highest error bars are obtained from the pure pseudo-$C_\ell$ estimator. The error bars from the minimum variance, quadratic estimator lie between those two. At the largest accessible scales, $2\leq\ell\leq100$, the error bars from the pure pseudospectrum estimator are $\sim1.25$ greater than the optimistic mode-counting estimation. Similarly, the error bars from the pure pseudospectrum estimators are at most $\sim1.1$ higher than the ones derived from the minimum variance, quadratic estimators. At the smaller angular scales where lensing dominates, the three approaches lead to almost the same uncertainties.\\

The behaviour of the uncertainties as a function of the sky fraction is common to the three approaches. At the smaller angular scales first (for multipoles greater than $\sim100$), the behavior is monotonic since the uncertainties systematically increase with the value of $f_\mathrm{sky}$. This is because at these scales, the variance is dominated by the noise, which increases with the sky fraction. At larger scales however (for multipoles smaller than $\sim100$), the uncertainties have a more intricate behaviour. First one notes that the uncertainties {\it decrease} from $f_\mathrm{sky}=1\%$ to $f_\mathrm{sky}=3.5\%$. This is because the variance is dominated by sampling variance, which decreases for higher values of $f_\mathrm{sky}$. Second, one notes that uncertainties at $\ell<100$ then {\it increases} for a sky coverage ranging from $3.5\%$ to 10\%. This means that for $f_\mathrm{sky}>3.5\%$, the noise is now dominating the variance. Once this transition value of $f_\mathrm{sky}\sim3.5\%$ is crossed, the noise contribution dominates the variance for all our considered angular scales, $\ell\in[20,1020]$. Therefore, the variance will monotonically increase with $f_\mathrm{sky}$ at all the relevant angular scales once $f_\mathrm{sky}>3.5\%$. (Note that an identical behaviour is observed for the other values of $r$, though the specific value of $f_\mathrm{sky}$ at which the transition occurs depends on the specific value of $r$.)

\subsection{Signal-to-noise ratio on $r$}
\subsubsection{Numerical results}
The signal-to-noise ratio on $r$ is computing using Eq. (\ref{eq:fisher}) considering the three above-described methods to estimate the uncertainties on the $B$-mode reconstruction, $\mathbf{\Sigma}_{bb'}$. We remind that the summation in (\ref{eq:fisher}) is performed over bandpowers with a bandwidth of $\Delta_b=40$ and considering a range of multipoles from $\ell=20$ to $\ell=1020$.

\begin{figure}
\begin{center}
	\includegraphics[scale=0.35]{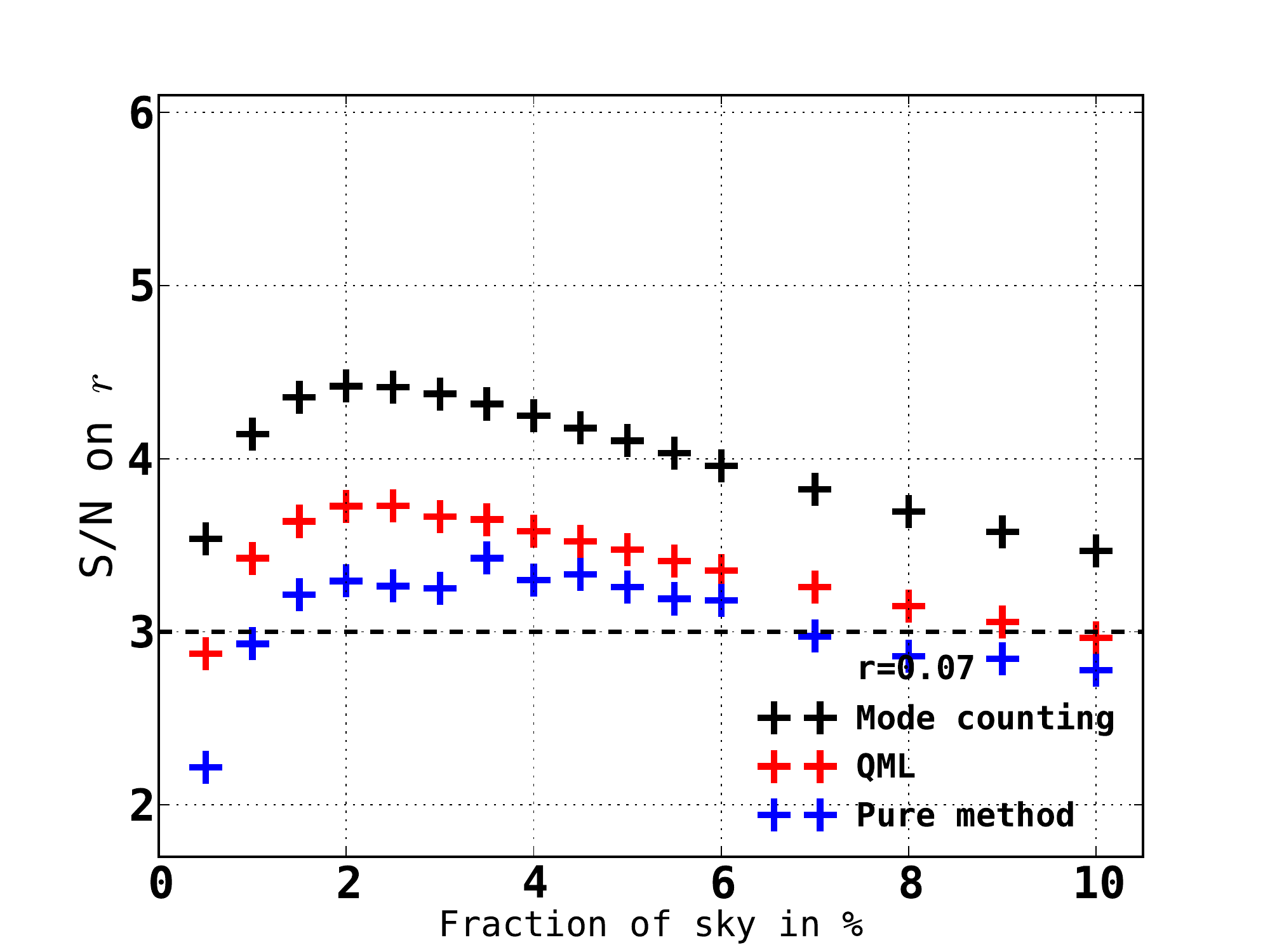}\vskip -0.1truecm
	\includegraphics[scale=0.35]{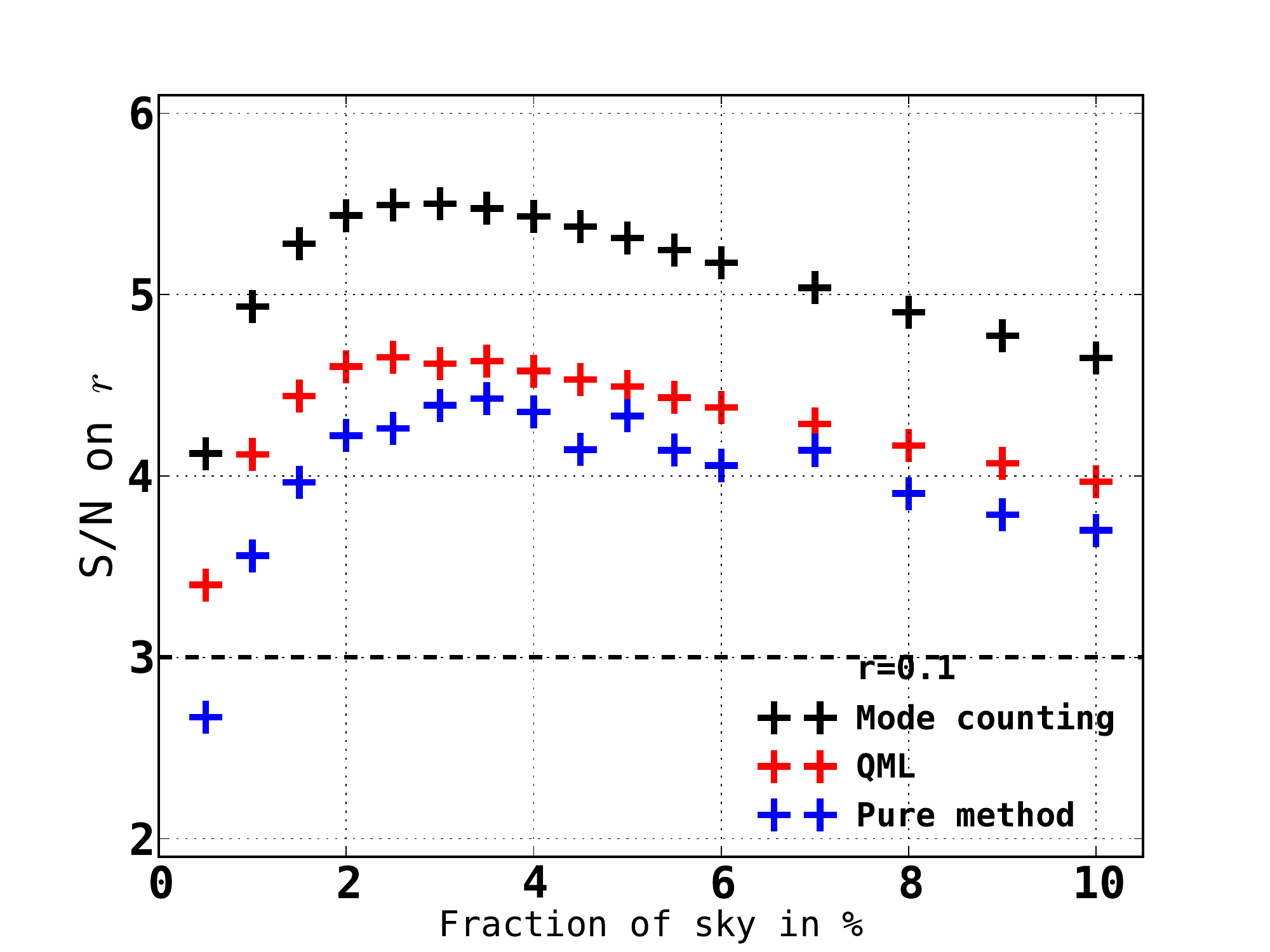}\vskip -0.1truecm
	\includegraphics[scale=0.35]{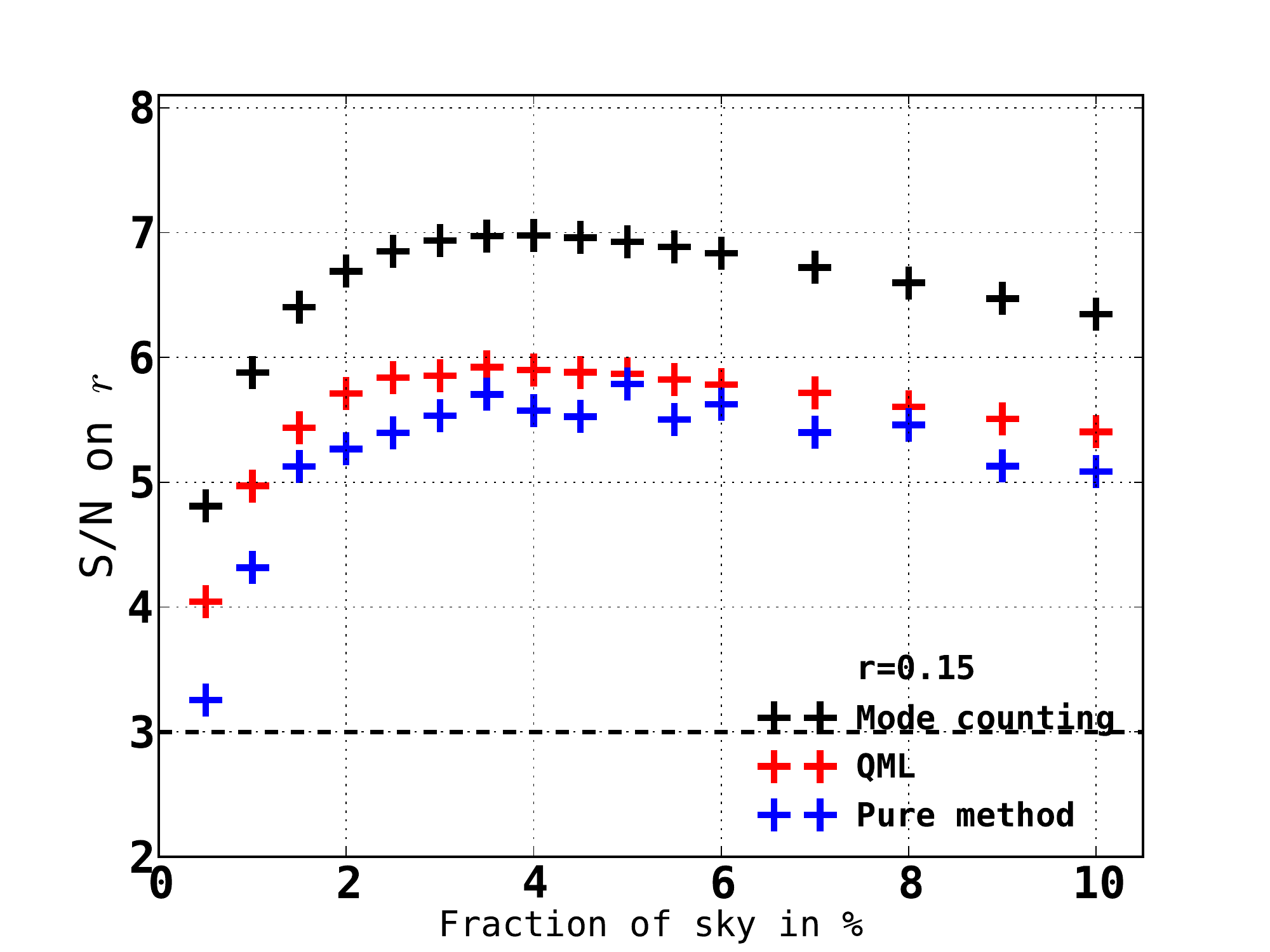}\vskip -0.1truecm
	\includegraphics[scale=0.35]{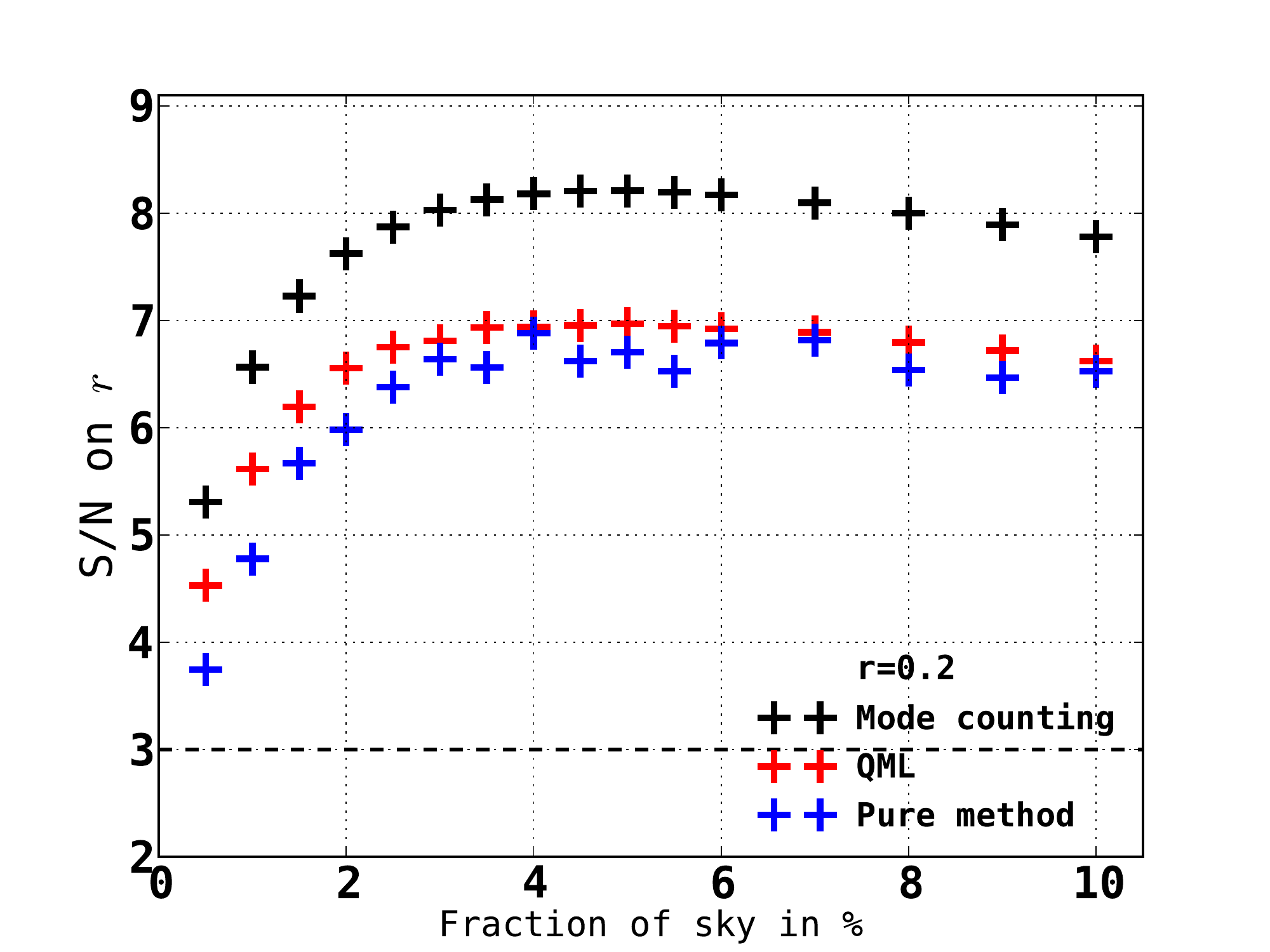}
	\caption{Signal-to-noise ratio for the estimation of $r$ from $B$-mode polarization data shown as a function of the sky coverage. The uncertainties are computed using three different approaches: mode-counting (black crosses), minimum variance quadratic estimator (red crosses) and pure pseudospectrum estimator (blue crosses). Each panel corresponds to a different fiducial value of the tensor-to-scalar ratio, $r=0.07,~0.1,~0.15$ and $0.2$ from top to bottom.}
	\label{fig:snrfsky}
\end{center}
\end{figure}

Our numerical results on the signal-to-noise ratio for $r$ are gathered in Fig. \ref{fig:snrfsky}, showing $\left(\mathrm{S/N}\right)_r$ as a function of the sky coverage. 
Each panel corresponds to a given value of the tensor-to-scalar ratio, $r=0.07,~0.1,~0.15$, and, $0.2$ (from top to bottom). 
For each panel, the black, red, and blue crosses correspond to the signal-to-noise ratio derived by using the mode-counting, the minimum variance quadratic estimator, and, the pure pseudo-$C_\ell$ estimator, respectively. 
The horizontal, dashed line marks a $3\sigma$ detection. The sky fraction varies from 0.1\% to 10\%, what is wide enough to sample the maximal values of the signal-to-noise ratio. 
We note that the signal-to-noise ratio keeps decreasing for $f_\mathrm{sky}>10\%$. 
This is because for the level of noise and values of $r$ here considered, the uncertainties on the reconstructed $B$-mode are noise dominated at all scales for $f_\mathrm{sky}>10\%$. 
Similarly, the (S/N)$_r$ keeps decreasing for $f_\mathrm{sky}<0.5\%$, because the uncertainties on angular scales greater than a degree are dominated by the sampling variance for such low values of the sky fraction. \\

For the case of the mode-counting first, the signal-to-noise ratio is systematically greater than 3 for all the considered values of the sky coverage, and for all the considered values of the tensor-to-scalar ratio.

Considering then the case of the minimum variance, quadratic estimator, the signal-to-noise ratio on $r$ is systematically greater than 3 for $r=0.1,~0.15$ and $0.2$. For a tensor-to-scalr ratio of $r=0.07$, the (S/N)$_r$ is greater or equal to three for $1\%\leq f_\mathrm{sky}\leq9\%$. 

Assuming finally a pure pseudospectrum reconstruction of the $B$-mode, the signal-to-noise ratio is systematically greater than 3 for $r=0.15$ and $r=0.2$ only. For a tensor-to-scalar ratio of $r=0.1$, a measurement of it with a (S/N)$_r$ of at least 3, is possible for $f_\mathrm{sky}\geq1\%$. For a smaller value of $r=0.07$, its measurement with  (S/N)$_r\geq3$ is possible assuming $1.5\%\leq f_\mathrm{sky}\leq7\%$. We note however here that for $r=0.07$ and $r=0.1$, the signal-to-noise ratios remains greater than 2 assuming a pure pseudospectrum reconstruction of $C^B_\ell$.\\

As expected from the error bars on the reconstructed $B$-mode, the highest and lowest (S/N)$_r$'s are respectively obtained from the mode-counting estimation, and the pure pseudospectrum estimator, while the (S/N)$_r$ from the minimum variance quadratic estimator lies between those two. This is the case for all the values of the tensor-to-scalar ratio we consider. At the peak, the signal-to-noise ratio from the pure pseudospectrum estimation of the $C^B_\ell$ is $\sim$15\% ($r=0.2$) to $\sim$20\% ($r=0.07$) smaller than the signal-to-noise ratio derived from the optimistic mode-counting. This means that the statistical significance on the measurement of $r$ by using the optimistic mode-counting is overestimated by a factor $\sim1.25$ as compared to the more realistic case of the pure pseudoreconstruction of the $B$-mode.

Similarly, the (S/N)$_r$ from the pure pseudospectrum estimator is $\sim$1.5\% ($r=0.2$) to $\sim$8\% ($r=0.07$) smaller than the signal-to-noise ratio derived from the minimum variance, quadratic estimator. Using the minimum variance, quadratic estimator to estimate the $B$-mode, as compared to the use of the pure pseudospectrum, thus translates into a gain  in the statistical significance on the measurement of $r$, of a factor 1.01 to 1.08. This gain appears rather small but is larger for smaller values of the tensor-to-scalar ratio.

\subsubsection{Optimization of the sky coverage}
As clearly shown in Figs. \ref{fig:snrfsky}, there exists a value of the sky coverage, which maximizes the signal-to-noise ratio on $r$. This {\it optimal} value of $f_\mathrm{sky}$ was already observed in Ref. \cite{jaffe_etal_2000}, using only the mode-counting expression for the statistical error bars on the $B$-mode estimation though. We found that such an optimal value also exists using the minimum variance quadratic estimator or the pure pseudo-$C_\ell$ estimator. This is intuitively understood as follows. The {\it statistical} uncertainties on the angular power spectrum estimation have two sources, the sampling variance, which is dominant at the largest angular scales, and the noise variance dominating at the smallest angular scales. Reducing the sampling variance is obtained by covering a large fraction of the sky. However, for a given sensitivity and a given time of observation, covering a large fraction of the sky inevitably translates into a higher level of noise per pixel. One should therefore find the good balance between sampling and noise variance so as to minimize the total error on given targetted parameters, which is $r$ here.

The salient features of those results are summarized in the table \ref{tab:fsky}. For each value of the tensor-to-scalar ratio and for each techniques used to compute uncertainties on the $B$-mode, we provide the values of the sky fraction maximizing the signal-to-noise ratio, $f^{\mathrm{(opt)}}_\mathrm{sky}$. Its associated (thus maximal) value of the signal-to-noise ratio, (S/N)$^{\mathrm{(opt)}}_r$ is also reported in this table. We stress that the position of the peak of (S/N)$_r$ is well defined for the mode-counting and the minimum variance, quadratic estimator. Such a position of the peaking signal-to-noise ratio is however less pronounced for the case of the pure pseudo-$C_\ell$ estimation of the $B$-mode (see e.g. the case $r=0.7$ for which a range of $2\%\lesssim f_\mathrm{sky}\lesssim6\%$ leads roughly to the same (S/N)$_r$). This means that the values of $f^{\mathrm{(opt)}}_\mathrm{sky}$ reported in Tab. \ref{tab:fsky} for the case of the pure pseudospectrum approach are more indicative than a sharply defined value. \\
\begin{table}
\begin{center}
\begin{tabular}{lm{0.5cm}m{0.75cm}m{0.75cm}m{0.75cm}m{0.75cm}} \hline\hline
	$r$ && 0.07 & 0.1 & 0.15 & 0.2 \\ \hline
	$f^{\mathrm{(opt)}}_\mathrm{sky}~[\%]$: && & & & \\
	\hspace*{0.35cm}Mode-counting && 2.0 & 3.0 & 4.0 & 5.0 \\
	\hspace*{0.35cm}Minimum-variance $C_\ell$ &&  2.5 & 2.5 & 3.5 & 5.0 \\
	\hspace*{0.35cm}Pure pseudo-$C_\ell$ &&  3.5 & 3.5 & 5.0 & 4.0  \\  \hline
	(S/N)$^{\mathrm{(opt)}}_r$: && & & & \\
	\hspace*{0.35cm}Mode-counting && 4.4 & 5.5 & 7.0 & 8.2  \\
	\hspace*{0.35cm}Minimum-variance $C_\ell$ &&  3.7 & 4.7 & 5.9 & 7.0  \\
	\hspace*{0.35cm}Pure pseudo-$C_\ell$ &&  3.4 & 4.4 & 5.8 & 6.9 \\ \hline\hline
\end{tabular}
	\caption{Values of the sky fraction maximizing the signal-to-noise ratio on $r$, $f^{\mathrm{(opt)}}_\mathrm{sky}$, and maximal values of the signal-to-noise ratio, (S/N)$^{\mathrm{(opt)}}_r$. This is given for each techniques used to estimate the uncertainties on the reconstruction of the $B$-mode. We notice that concerning the pure method, the position of the maximum value is not as defined as the other methods (see text).
	%\begin{color}{red} For pure pseudospectrum, the SNR is rather flat in some values of the sky coverage. Shall we want to also provide a kind of range of optimal sky fraction (and its associated range of SNR) ? (cf. the text)\end{color}
	}
	\label{tab:fsky}
\end{center}
\end{table}	

For all the approaches used to estimate the uncertainties on the $B$-mode, we observe that the optimal sky fraction increases with the value of the tensor-to-scalar ratio. This is because for higher values of $r$, the signal in the $B$-mode is higher. One should therefore minimize first the sampling variance by increasing the observed part of the sky. \\

Except for the case of $r=0.2$, we note that the optimal sky coverage assuming a minimum variance, quadratic estimator slightly differs by 0.5\% (either higher or lower) than the value of $f^\mathrm{(opt)}_\mathrm{sky}$ (in \%) as inferred from the mode-counting. We also note that the optimal sky fraction obtained for the pure pseudo-$C_\ell$ reconstruction of the $B$-mode differs by 1\% to 1.5\% (depending on the value of $r$) from the one inferred from the mode-counting estimation of the uncertainties on the $B$-mode. Nevertheless, the values of the sky fraction for which the detection of the tensor-to-scalar ratio is peaking in the case of the mode-counting expression and the minimum-variance quadratic estimator fall in the range of optimized $f_\mathrm{sky}$ as derived from the pure pseudo-$C_\ell$ estimator. Those numerical results therefore show that (at least) for the range of values of $r$ here-considered, the value of the sky coverage, which maximizes the measurement of the tensor-to-scalar ratio is rather independent on the adopted method for evaluating the statistical uncertainties on the $B$-mode reconstruction. This means that using the mode-counting expression, though underestimating the error bars, allows for a rapid and reliable search of the range of values of the optimized sky fraction. (Obviously, such an optimization of $f_\mathrm{sky}$ based on the mode-counting expression is reliable providing the final data set to be analyzed using either the minimum-variance quadratic estimator or the pure pseudospectrum estimator.)

\subsubsection{Impact of binning}
For the two specific cases of the mode-counting uncertainties and the minimum variance, quadratic estimator, we note that an explicit reconstruction of the power spectrum is not mandatory to derive the (S/N)$_r$ in the Fisher formalism. 
One can indeed directly plugged in Eq. (\ref{eq:fisher}) the formulas (\ref{eq:mcountcov}) or (\ref{eq:qmlcov}). 
This allows for a study of the impact of binning on the signal-to-noise ratio, letting the bandwidth to vary from $\Delta_b=1$ (i.e. no binning) to $\Delta_b=40$ (i.e. the binning imposed by the use of the pseudospectrum estimator in this analysis).
\begin{figure}
\begin{center}
	\includegraphics[scale=0.4]{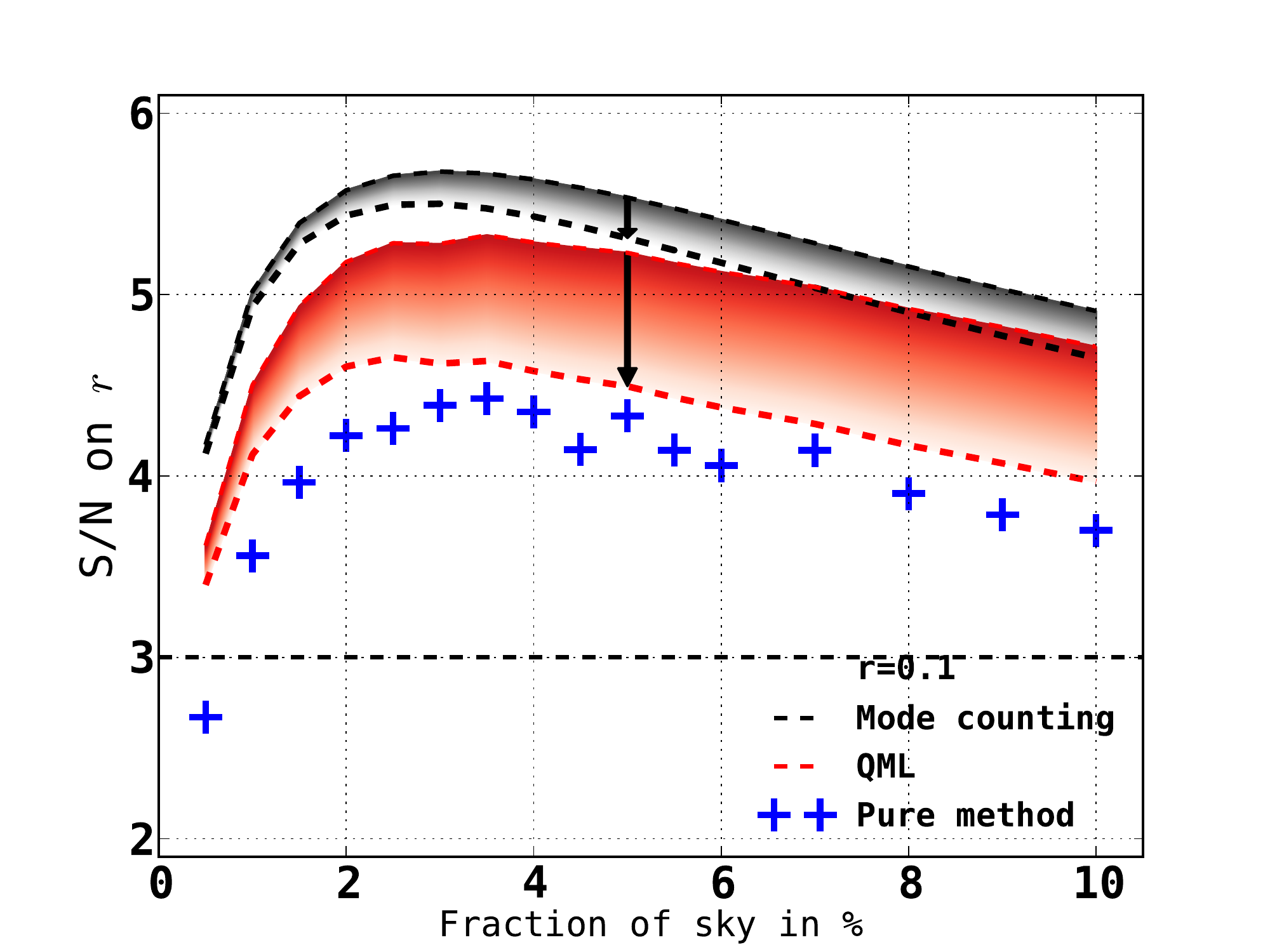}
	\caption{Signal-to-noise ratio for $r=0.1$ as a function of the observed sky fraction, derived for three methods used to estimate the uncertainties on the $B$-mode reconstruction: mode-counting (black area), minimum variance quadratic estimator (red area), and, pure pseudospectrum estimator (blue crosses). For the two first methods, we let the bandwidth of the bins to vary from $\Delta_b=1$ (highest (S/N)$_r$) to $\Delta_b=40$ (lowest (S/N)$_r$). For the specific case of the pure pseudospectrum estimator, the reconstruction of the $C^B_\ell$ requires to use the bandwidth $\Delta_b=40$. (We remind that the range of multipoles used to compute the signal-to-noise ratio is $20\leq\ell\leq1020$.)}
	\label{fig:snrbin}
\end{center}
\end{figure}

The impact of binning is illustrated in Fig. \ref{fig:snrbin} showing the signal-to-noise ratio on $r=0.1$ as a function of the sky coverage. The grey (red) area corresponds to the (S/N)$_r$ using the mode-counting (minimum variance quadratic estimator) to estimate the uncertainties the angular power spectrum of the $B$-mode. For each shaded area, the highest signal-to-noise ratio is obtained for $\Delta_b=1$ and the lowest for $\Delta_b=40$. As a reference, we also show the (S/N)$_r$ obtained with pure pseudospectrum reconstruction (thus using a bandwidth of $\Delta_b=40$) depicted by the blue crosses. The overall effect of increasing the width of the bandpower is to lower the signal-to-noise ratio. The decrease is however more pronounced for the case of the minimum variance, quadratic estimator than for the mode-counting estimation of the error bars on the reconstructed $B$-mode. This is due to the fact that correlations between multipoles (or bandpowers) are accounted for in the minimum variance, quadratic estimator, while those are supposed to be systematically vanishing for the mode-counting estimation of the covariance matrix. This additional piece of information contained in the correlations is therefore partially lost by averaging over bandpowers.
We also checked that artificially imposing those off-diagonal correlations to be zero lowered the signal-to-noise ratio in the minimum variance method, although we note that once the bins are sufficiently wide the effect of the binwidth on the (S/N)$_r$ should be weak.

The maximum values of the (S/N)$_r$ obtained using a bandwidth of $\Delta_b=1$, and a bandwidth of $\Delta_b=40$, are reported in Tab. \ref{tab:bin}, for each values of $r$ and for the mode-counting and the minimum variance quadratic estimator. For each cases, we also report the value of the sky coverage corresponding to that maximum. For the mode-counting approach, increasing the bandwidth from $\Delta_b=1$ to $\Delta_b=40$, degrades the maximum (S/N)$_r$ by a factor $\sim9\%$ for $r=0.07$ and $0.1$, and, by a factor $\sim2\%$ for $r=0.15$ and $0.2$. This however only mildly affects the values of the sky fraction at which the maximum is achieved. The impact of binning is more marked for the minimum variance quadratic estimator however. Increasing the bandwidth from $\Delta_b=1$ to $\Delta_b=40$, here degrades the maximum (S/N)$_r$ by a factor $\sim11\%$ for all the values of the tensor-to-scalar ratio considered in this study. Similarly, the values of the sky fraction (in \%) at which this maximum is achieved is systematically lowered (except for the case $r=0.7$), by 1\% for $r=0.1$ and by 2\% for $r=0.2$. We note that despite these changes in the value of $f^\mathrm{(opt)}_\mathrm{sky}$ with the bandwidth, the optimized values of the sky fraction still fall in the range of optimized $f_\mathrm{sky}$ as derived from the pure pseudo-$C_\ell$ estimator.
\begin{table*}
\begin{center}
\begin{tabular}{lm{1.cm}m{2.cm}m{2.cm}m{2.cm}m{2.cm}} \hline\hline
	$r$ && 0.07 & 0.1 & 0.15 & 0.2 \\ \hline
	(S/N)$^{\mathrm{(opt)}}_r$ and $f^{\mathrm{(opt)}}_\mathrm{sky}$: && & & &  \\
	%&& & & &  \\
	mode-counting: && & & & \\
	\hspace*{0.35cm}$\Delta_b=1$ &&  4.6~(2.5\%) & 5.7~(3\%) & 7.2~(4\%) & 8.4~(5.5\%)  \\
	\hspace*{0.35cm}$\Delta_b=40$ &&  4.4~(2.0\%) & 5.5~(3\%) & 7.0~(4\%) & 8.2~(5\%) \\  \hline
	minimum-variance $C_\ell$: && & & & \\
	\hspace*{0.35cm}$\Delta_b=1$ &&  4.2~(2.5\%) & 5.3~(3.5\%) & 6.8~(5\%) & 8.0~(7\%)  \\
	\hspace*{0.35cm}$\Delta_b=40$ &&  3.7~(2.5\%) & 4.7~(2.5\%) & 5.9~(3.5\%) & 7.0~(5\%)  \\ \hline\hline
\end{tabular}
	\caption{Maximum values of the (S/N)$_r$ for the mode-counting method (upper part), and the minimum variance quadratic estimator (lower part). This is derived assuming no-binning (i.e. $\Delta_b=1$), or using a binning with a bandwidth of $\Delta_b=40$. In parentheses is the values of $f_\mathrm{sky}$ (in percent) at which this maximum is reached.}
	\label{tab:bin}
\end{center}
\end{table*}

%%%%%%%%%%%%%%%%%%%%%%%%%%%%%%%%%%%%%%%%%%%%%%%%%
\section{Measuring the tensor-to-scalar ratio: selected examples}
\label{sec:realistic}
\subsection{Experimental setups}
We turn to the question of the detection of $r$ in more realistic cases. Clearly, a spherical cap is ideal. The issue of leakages is strongly related to the detailed shape of the contours of the observed (or kept-in-the-analysis) portion of the sky (see e.g. the figure 20 of Ref. \cite{grain_etal_2009} for the impact of the shape of the mask on the statistical error bars). A spherical cap then leads to the smallest amount of leakages for a given sky fraction since its contour has the smallest perimeter for that given sky fraction. To this end, we consider three archetypal cases, which capture the main characteristics of ongoing, or being-deployed, small-scale experiments (ground-based or balloon-borne), a possible upgrade of those ground-based experiments to an array covering a rather large fraction of the sky ($\sim50\%$), and,  a possible satellite mission covering the entire celestial sphere. \\
\begin{figure*}
\begin{center}
\begin{tabular}{ccc}
	\includegraphics[scale=0.25]{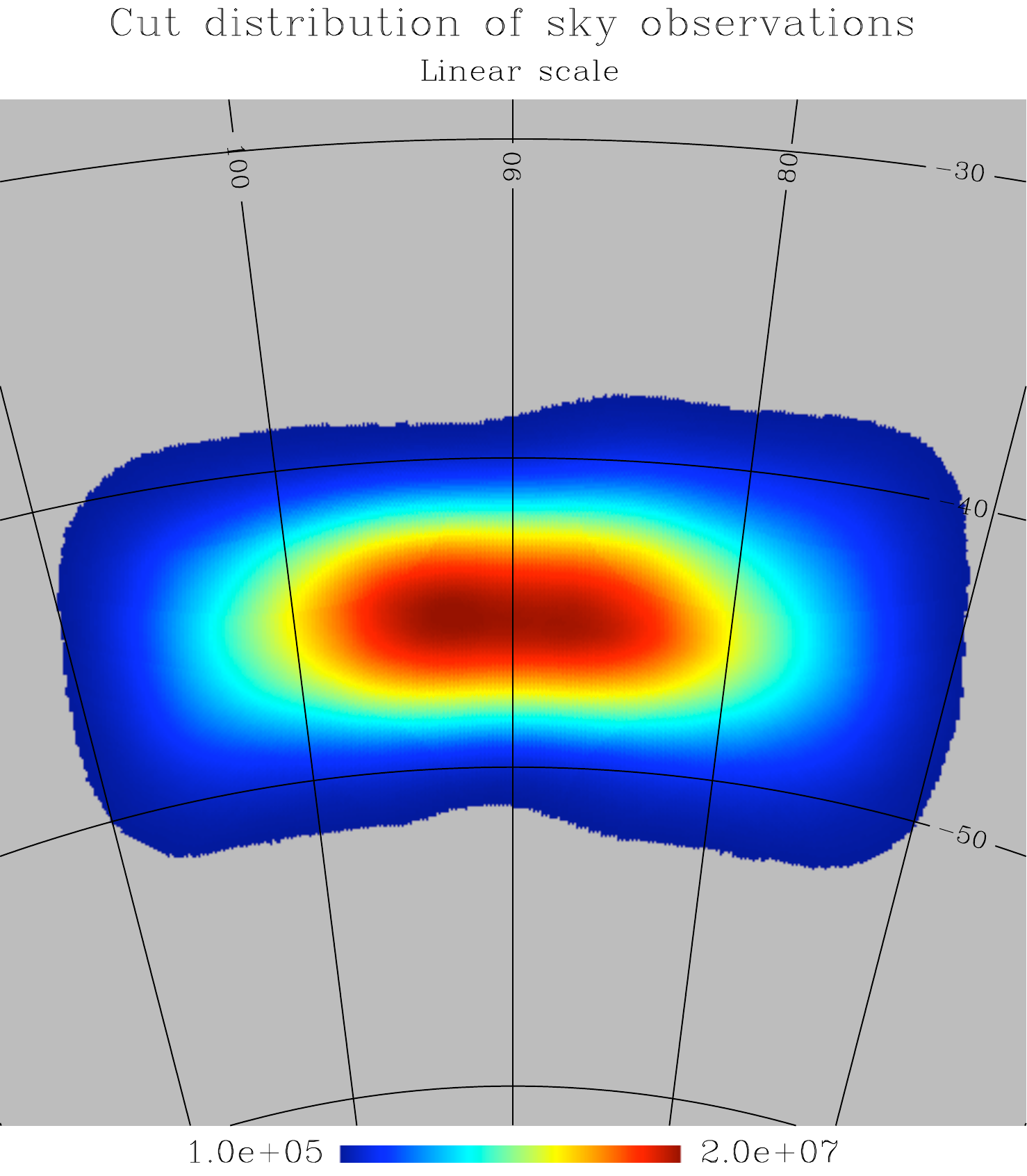}&\includegraphics[scale=0.25]{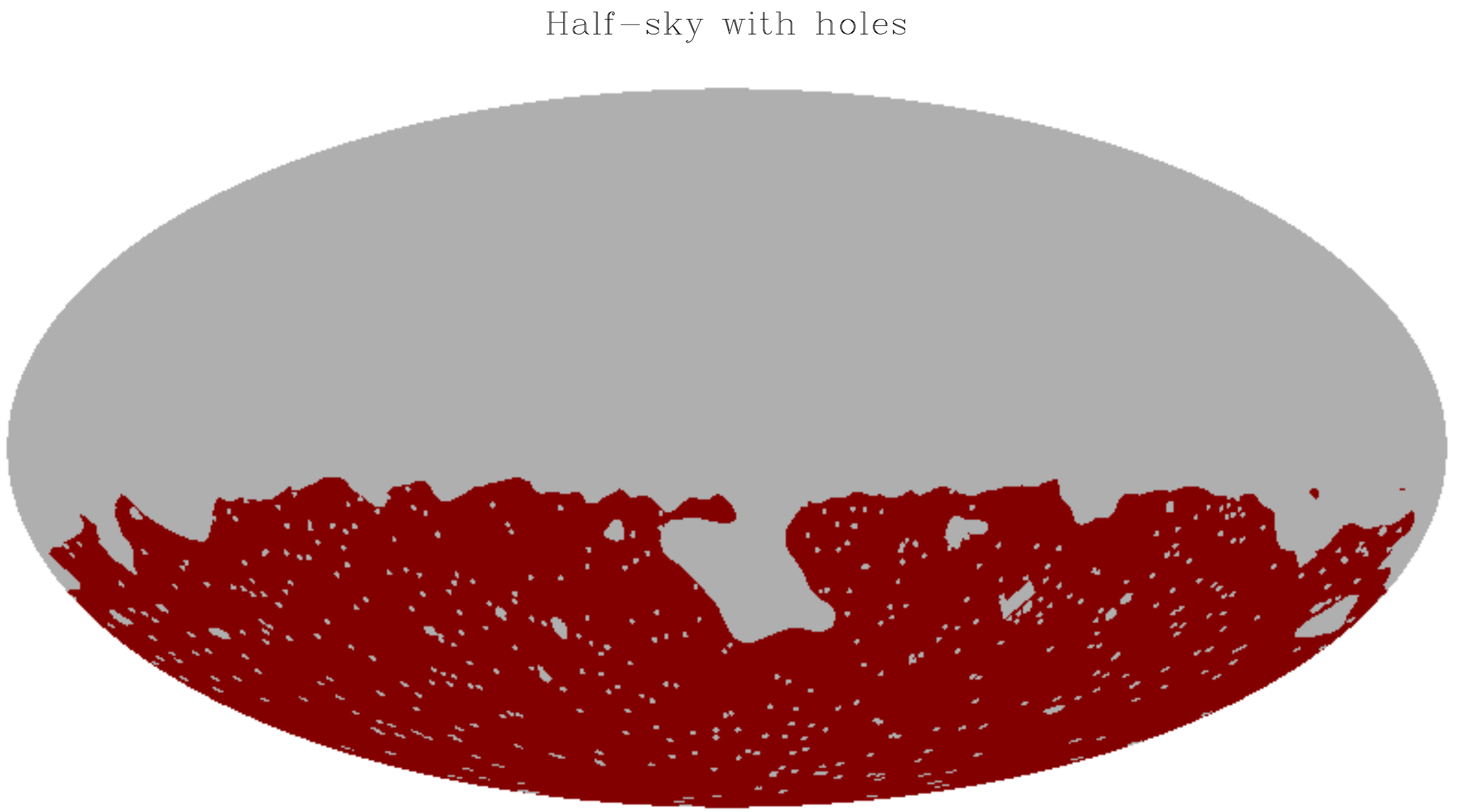}&\includegraphics[scale=0.25]{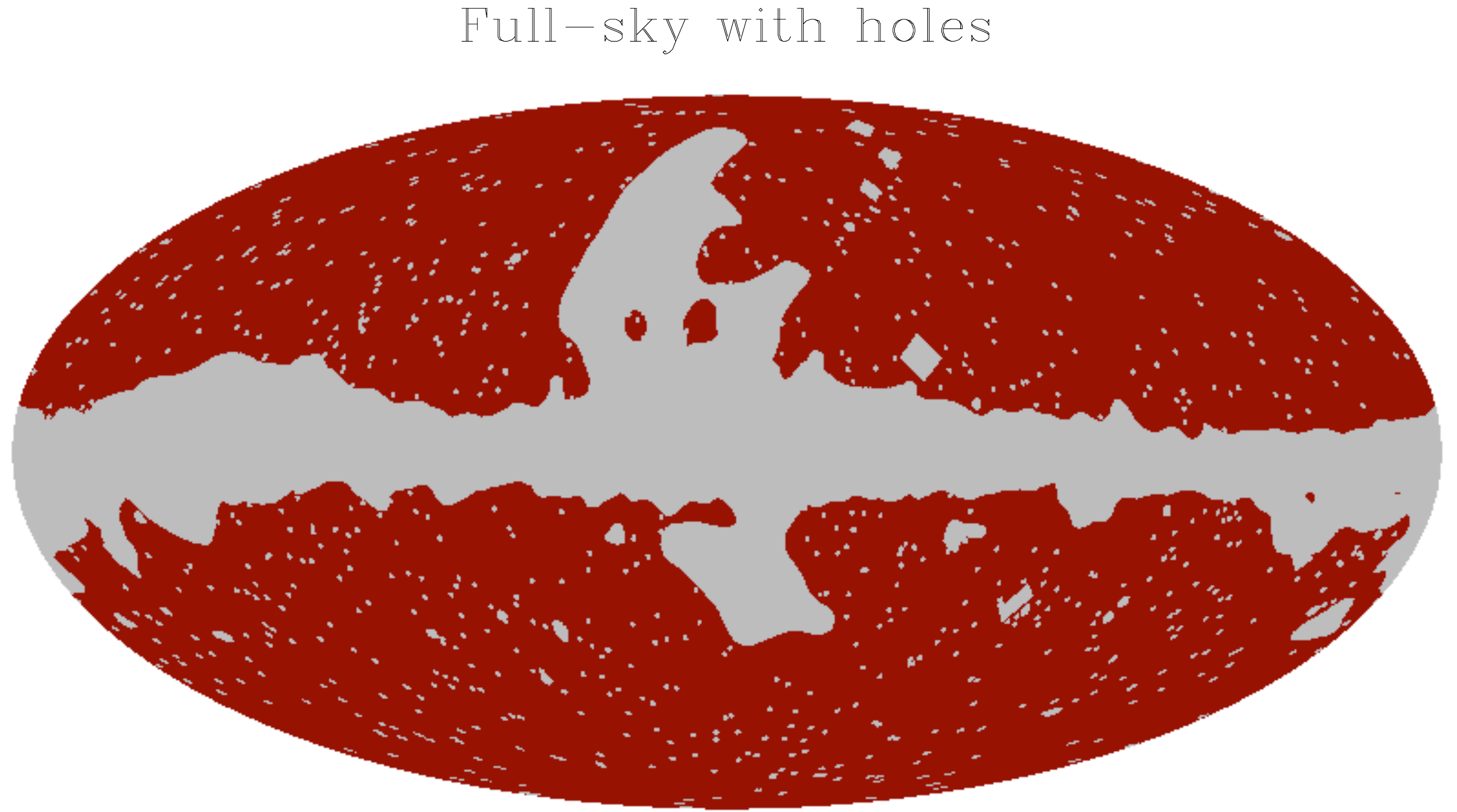}
\end{tabular}
	\caption{{\it Left Panel:} A potential noise distribution (number of hits per pixel) for small-scale experiments covering $\sim$1\% of the sky. The noisiest pixels have been discarded and the noise distribution ranges from $10^5$ to $10^7$. {\it Middle panel:} Portion of the sky kept for estimating the $B$-mode angular power spectra for a possible array of ground-based telescopes. It roughly covers 36\% of the sky (Note that for this analysis the choice of galactic or ecliptic south hemisphere is equivalent, see text for more information). {\it Right panel:} Portion of the sky kept for estimating the $B$-mode angular power spectra for a possible satellite mission, roughly covering 71\% of the sky.}
	\label{fig:skycoverage}
\end{center}
\end{figure*}

First, we consider a part of the sky, which is typical of small-scale experiments, inspired by the design study of the {\sc ebex} experiment \cite{britt_etal_2010} and very similar to the patch and noise level used in Ref. \cite{louis_etal_2013} mimicking current ground-based efforts such as {\sc ACTpol} or \pb. This roughly covers 1\% of the sky with a highly inhomogeneous noise, as shown in the left panel of Fig. \ref{fig:skycoverage}. The average noise level is $\sim5.75\mu$K-arcminute and the beamwidth is 8 arcminutes. This setup will be referred as "small-scale experiment" hereafter.

Second, we consider the potential case of an array of ground-based telescopes allowing for covering $\sim50\%$ of the celestial sphere with a (here-considered homogeneous) noise of $1\mu$K-arcminute and a beamwidth of 3 arcminutes. These specifications roughly correspond to the Stage IV of future CMB experiment as reported in e.g. Ref. \cite{snowmass}. For such a large fraction of the sky, masking the regions with high foreground galactic emissions is needed. To this end, we consider that the entire galactic south hemisphere would be observed and we apply a galactic mask, and a mask for point sources. We make use of the {\it R9} galactic mask used for polarized data of {\sc wmap} and add the point-sources mask \cite{wmapmask}. The resulting sky coverage is depicted in the middle panel of Fig. \ref{fig:skycoverage}, and it roughly covers $\sim$36\% of the sky. (We note that this sky coverage is just the restriction to the galactic south hemisphere of the kept-in-the-analysis portion of the sky for a satellite mission as defined below.) This configuration will be referred as "array of telescopes" in the following. One could have chosen also the south ecliptic hemisphere, however while this consideration is important for e.g. scanning strategy and foreground treatment, this is less relevant for the minimization of the $E$-to-$B$ leakage for which one key point is the complexity of the contours of the patch, which would be reflected in both choices of observation. 

Third, we consider a potential satellite-like mission with homogeneous noise at a level of $2.2\mu$K-arcminute and a beamwidth of 8 arcminutes \cite{bock_etal_2008}. The portion of the sky to be analysed is displayed in the right panel of Fig. \ref{fig:skycoverage}. It corresponds to $~71\%$ of the entire celestial sphere. As for the previous case, the removed portion of the sky corresponds to the {\sc wmap} galactic and point-sources mask. This will be referred as "satellite mission" hereafter.

\subsection{Sky apodizations}
For each of those cases, the uncertainties on the reconstructed angular power spectrum are obtained using the pure pseudospectrum estimator, using the same set of bandpowers as in Sec. \ref{ssec:powspec}. We note however that the first bandpower, ranging from $\ell=2$ to $\ell=20$ will be explictly shown hereafter since those scales are now accessible for the case of an array of telescopes, and the case of a satellite mission.

Optimizing the sky apodization to be applied to the maps is a key step to reach better performances on the reconstructed $C^{B}_\ell$. Two classes of sky apodizations have been proposed in the literature (see Refs. \cite{smith_zaldarriaga_2007,grain_etal_2009}). The first class consists in an analytical formula fullfilling the appropriate Neuman and Dirichlet boundary conditions using arches of a sine function. The apodization length can be subsequently optimized, bin-per-bin, to minimize the uncertainties on the estimated $C^{B}_\ell$. The second class is a set of minimum-variance optimized sky apodizations. They can be computed either in the pixel domain as originally proposed in \cite{smith_zaldarriaga_2007} (thus allowing for relaxing the boundary conditions and keeping track of the informations about the $B$-modes contained in the so-called ambiguous modes), or in the harmonic domain as proposed in \cite{grain_etal_2009} providing the noise is homogeneous but here forcing the Neuman and Dirichlet boundary conditions to hold. The pixel-based computation is more flexible and general (being e.g. applicable to cases of inhomogeneous noise), and it was shown to generically lead to better performances. It is however numerically costly, while the harmonic-based computation is very rapid\footnote{This is because those optimized sky apodization are obtained as the solution of linear system of size $N_\mathrm{pix}$. This linear system corresponds to a convolution, which is drastically simplified in the harmonic domain (the convolution kernels becoming diagonal) assuming the noise is homogeneous.}. \\

\begin{figure*}
\begin{center}
\begin{tabular}{ccc}
	\includegraphics[scale=0.25]{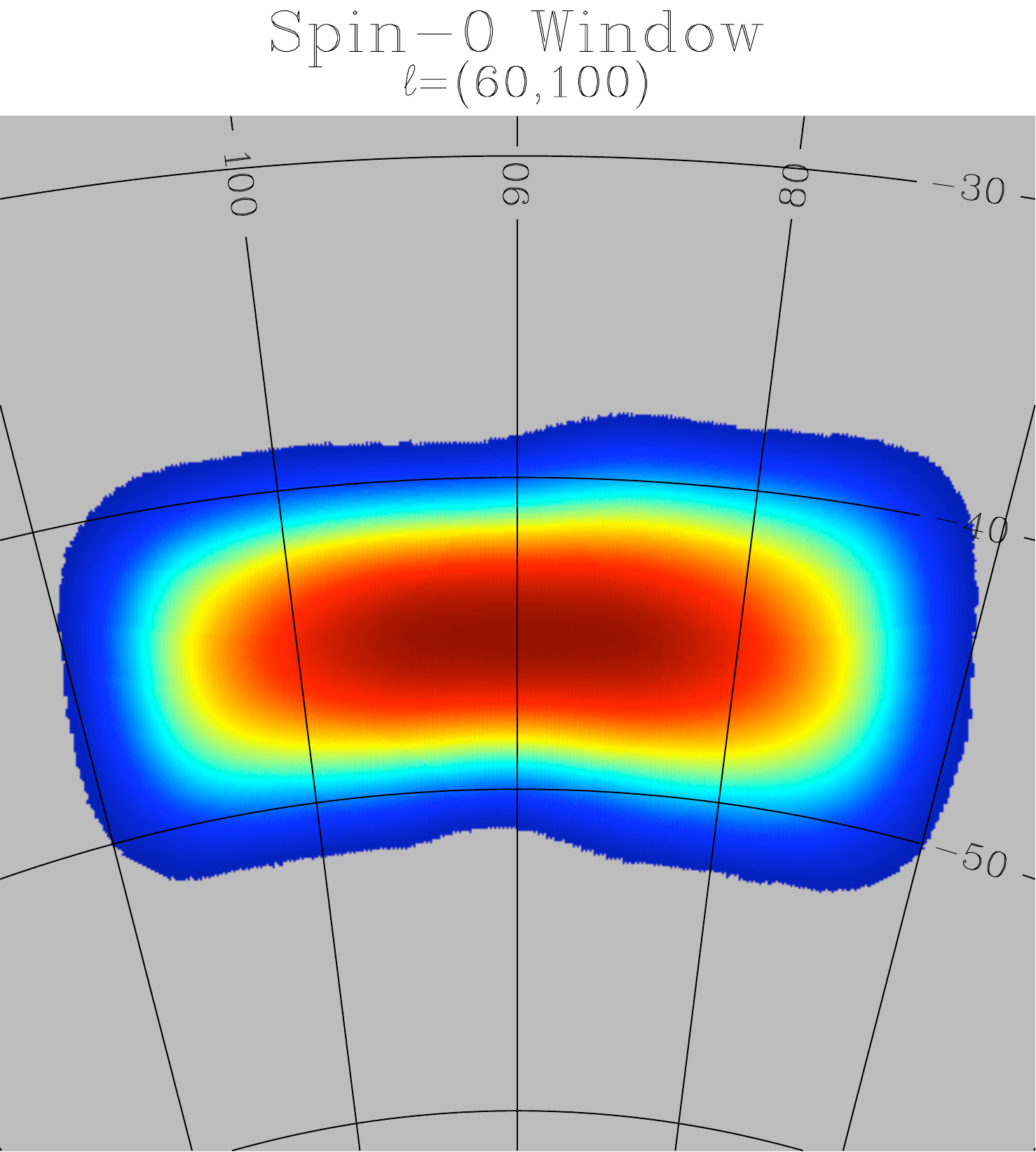}&\includegraphics[scale=0.25]{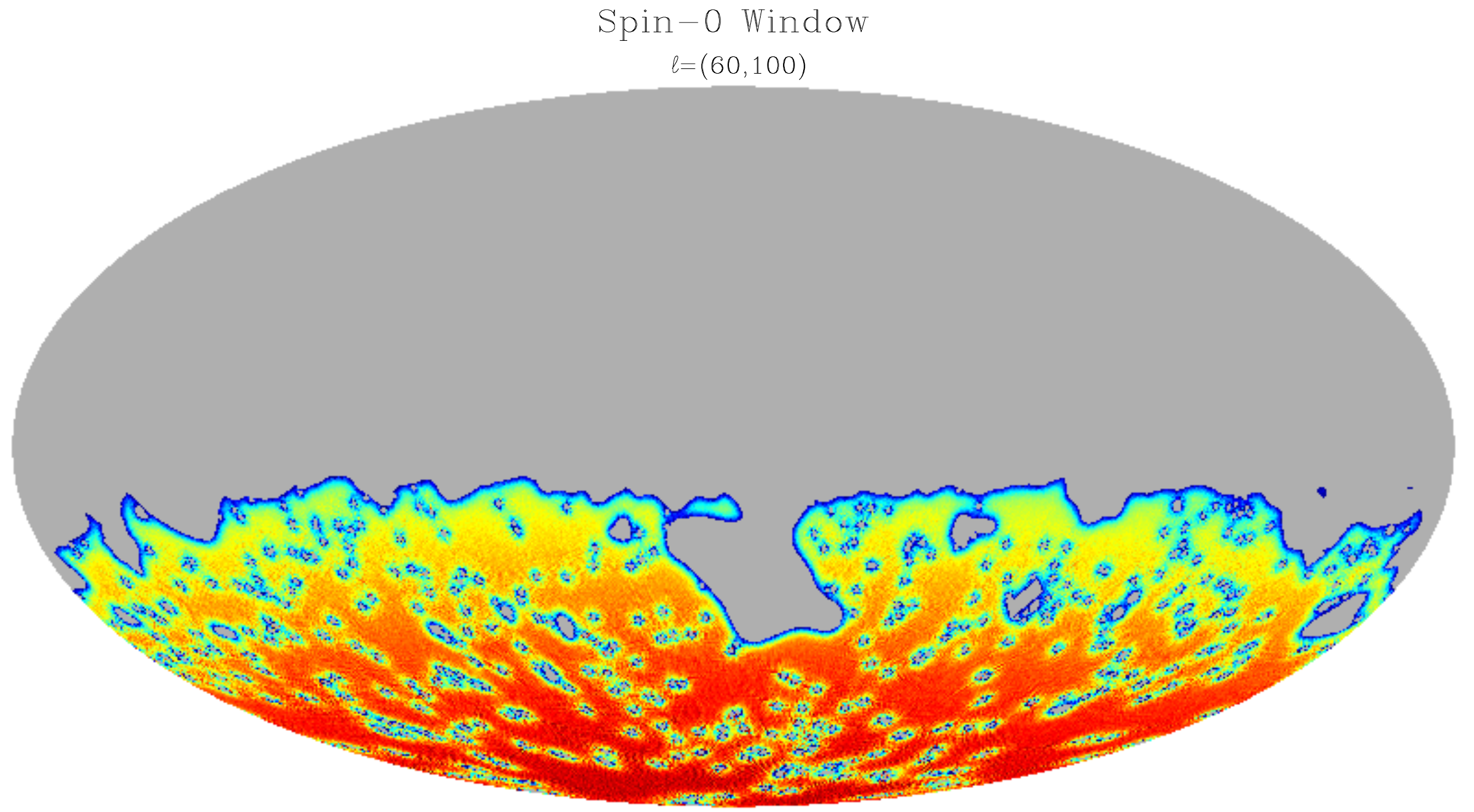}&\includegraphics[scale=0.25]{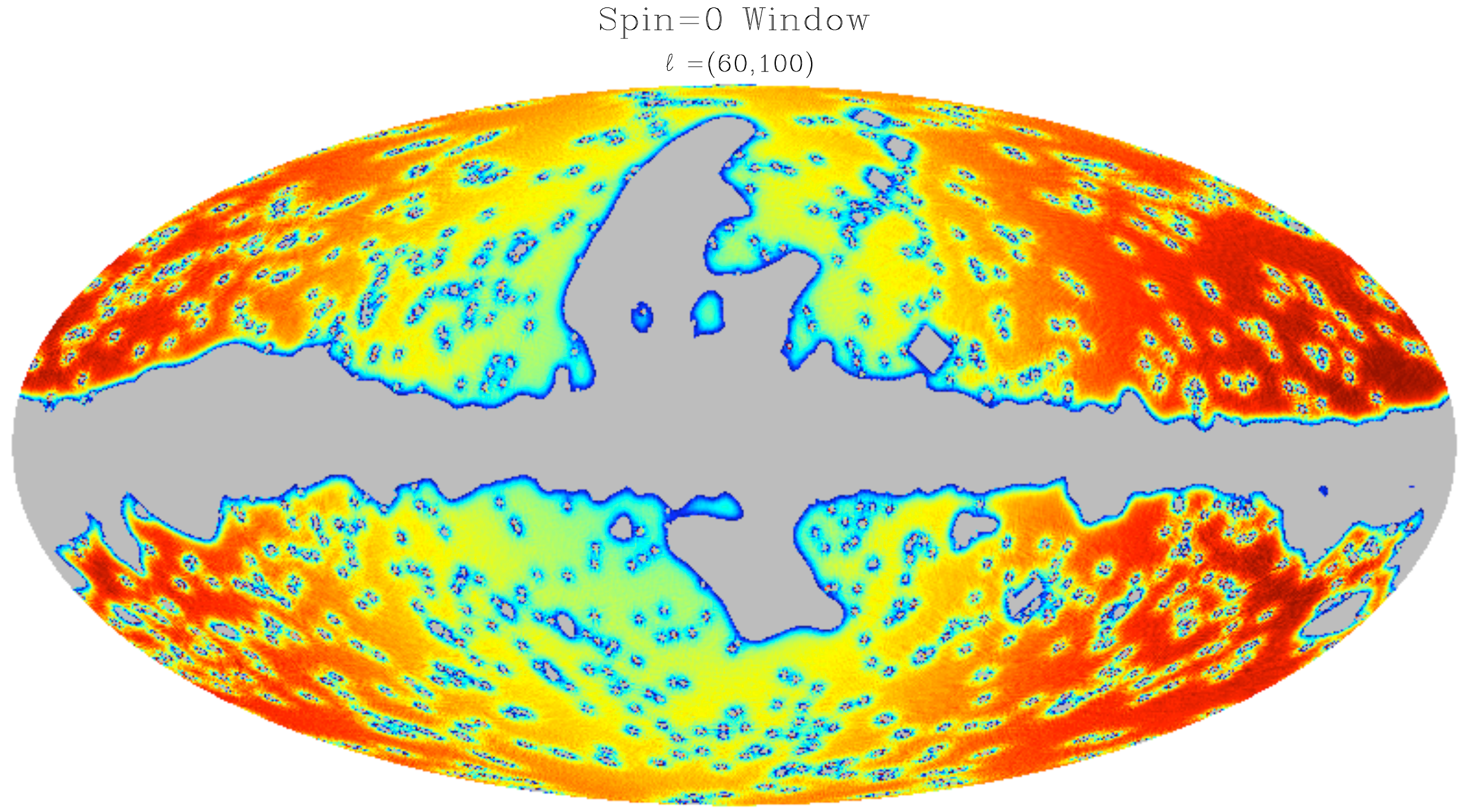}
\end{tabular}
	\caption{Spin-0 (scalar) component of the optimized sky apodization for a small-scale experiment covering $\sim$1\% of the sky with inhomogeneous noise (left panel), for a possible array of ground-based telescope covering $\sim$36\% of the sky with homogeneous noise (middle panel), and, for a possible satellite mission covering $\sim$71\% of the sky with homogeneous noise (right panel). This sky apodization is optimized for a bandpower ranging from $\ell=60$ to $\ell=100$ and for a value of the tensor-to-scalar ratio $r=0.05$.}
	\label{fig:skyap}
\end{center}
\end{figure*}

To illustrate the specific case of the minimum-variance sky apodizations (computed in the pixel domain), an example of its scalar, spin-0 component is depicted in Fig. \ref{fig:skyap} for the three experimental setups. This sky apodization has been optimized for the bandpower comprised between $\ell=60$ and $\ell=100$ and for a value of the tensor-to-scalar ratio $r=0.05$. We note that the sky apodization computed on the sole galactic south hemisphere for an array of telescopes (see middle panel of Fig. \ref{fig:skyap}) is {\it not} the restriction to the galactic south hemisphere of the sky apodization for a satellite mission computed on both hemispheres (see right panel of Fig. \ref{fig:skyap}). (This would obviously be the case for analytic sky apodizations.) This is mainly due to the fact that the global shape of the observed portion of the sky is taken into account in the optimization process. This is clearly seen by noticing that for a typical patch covering both hemispheres, there is a rather large apodization length parallel to the azimuthal direction while the apodization length along the zenithal direction is rather small. On the contrary for a typical patch covering only the south hemisphere, the apodization length is essentially along the zenithal direction. Such a difference is explained by the important hole along the zenithal direction in the north hemisphere, which leaves its footprint on the final shape of the sky apodization. (There is a second source of differences due to the different beamwidth and noise level, which are considered for both types of experiments. They however have a tiny impact on the orientation of the apodization length since they are evenly distributed over the sky and therefore do not have any preferred directions.) \\

A couple of comments about the numerical computation of the pixel-based, minimum-variance sky apodizations are in order here. They are theoretically built to give the smallest uncertainties in the context of the pure pseudospectrum estimators. However, they are pratically computed from a Preconditionned Conjugate Gradient (PCG) algorithm, which efficiency strongly depends on the experimental configurations, especially with respect to the noise level and its distribution over the patch, as well as with respect to the complexity of the contours of the mask. 

First, the number of iterations in our implemented PCG rapidly increases for lower levels of noise: at the largest angular scales ($\ell\leq20$) the number of iterations ranges from $\sim100$ for a noise level of 5.75$\mu$K-arcminute to $\sim3000$ for a noise level of 1$\mu$K-arminute (the number of iterations being one order of magnitude smaller for smaller angular scales, $\ell\geq20$). The $B$-mode angular power spectrum is estimated using the same binning as in the previous section, leading to $N_\mathrm{bin}=26$, and we selected 6 values of $r$. Considering 3 experimental setups, this would translate into $\sim$500 optimized sky apodization to compute, which is numerically too costly. We therefore compute the optimized sky apodization for $r=0.05$ only but use them for all the here-considered values of $r$, meaning that the signal-to-noise ratios obtained for $r\neq0.05$ may be suboptimal within the context of the pure pseudospectrum estimator\footnote{In the specific case of small-scale experiments, it was however shown in Ref. \cite{grain_etal_2009} that the optimization process is mainly driven by the noise level and the amount of $E$-modes leaking into the $B$-mode, and poorly affected by the amplitude of the primordial $B$-mode. This means that the resulting sky apodizations may be mildly dependant on the assumed value of $r$ and that the derived signal-to-noise ratios are only slightly suboptimal for the case of small-scale experiments.}. 

Second, it is not guaranteed that the algorithm converges towards the optimal solution, especially for inhomogeneous noise (see Sec. IV C of Ref. \cite{grain_etal_2009} where it was shown that trimming out the external, noisiest pixels is required), or a very low level of noise (see Ref. \cite{smith_zaldarriaga_2007} mentionning that convergence is not reached for a noise level of $\sim1\mu$K-arminute, corresponding to the level of the array of telescopes case). This means that the performances of those sky apodizations have to be assessed using numerical simulations at the level of power spectrum reconstruction, comparing the resulting error bars on the estimated power spectra to the error bars obtained by using the other types of sky apodizations.

%\subsection{Numerical results}
\subsection{Power spectrum uncertainties}
The relative performances of the different sky apodizations are appraised at the level of power spectrum uncertainties. For each case we performed a series of 500 Monte-Carlo simulations to compute the statistical uncertainties on the reconstructed $B$-mode angular power spectra, assuming different kinds of sky apodizations. Such performances have been exhaustively studied for the small-scale experiment case and the satellite mission case (see Refs. \cite{smith_zaldarriaga_2007,grain_etal_2009} and Ref. \cite{ferte_etal_2013}, resp.). On the contrary, the applicability of the pure pseudospectrum estimator for the case of an array of telescopes was hitherto not studied. We then performed numerical simulations using the different classes of sky apodizations to assess the efficiency of the pure pseudospectrum reconstruction of the $B$-mode, and subsequently select those sky apodizations, which lead to the smallest uncertainties. 

In this section, we only briefly review the major conclusions concerning the cases of a small-scale experiment and a satellite mission. Then, we present the results of our numerical investigations for the case of an array of telescopes.

\subsubsection{Optimal apodizations: small-scale experiments and satellites missions} 
For the small-scale experiment case, it was shown that the lowest uncertainties in the range $\ell\in[2,1020]$ were obtained using either the pixel-based optimized sky apodizations or analytic sky apodizations appropriately chosen to minimize the variance per each bin. However, the harmonic-based computation of the sky apodization fails in providing error bars comparable to the previous ones in the entire range of multipoles considered here, simply because the noise is inhomogeneous. (We refer to Fig. 24 of Ref. \cite{grain_etal_2009} and discussions therein.) 

For the satellite mission case, the pixel-based computation of the minimum-variance sky apodization yields the smallest uncertainties for the range $\ell\in[2,1020]$. Similar performance is obtained by using the harmonic-based computation of these sky apodizations for $150\lesssim\ell\lesssim600$. The error bars however drastically increase for larger angular scales, the reason for that being the intricate contours of the galactic mask and the point-sources mask, which require to relax the Neuman and Dirichlet boundary conditions to keep (part of) the informations about the $B$-mode contained in the ambiguous modes. For such a case, analytic sky apodizations fails in providing comparable error bars at the largest angular scales. (We refer to Fig. 4 of Ref. \cite{ferte_etal_2013} and discussions therein.)

\subsubsection{Optimal apodizations: arrays of telescopes} 
For the case of an array of telescopes, we systematically search for the type of sky apodizations, which lead to the smallest uncertainties bin per bin and for each values of $r$ considered in this study. We first found that the harmonic-based, optimized sky apodizations yield to error bars higher than the analytic sky apodizations or the pixel-based, optimized sky apodizations. This is similar to what was observed for the case of a satellite mission and the inefficiency of the harmonic-based, optimized sky apodizations is due to the complexity of the contours of the mask. 

\begin{figure}
\begin{center}
	\includegraphics[scale=0.5]{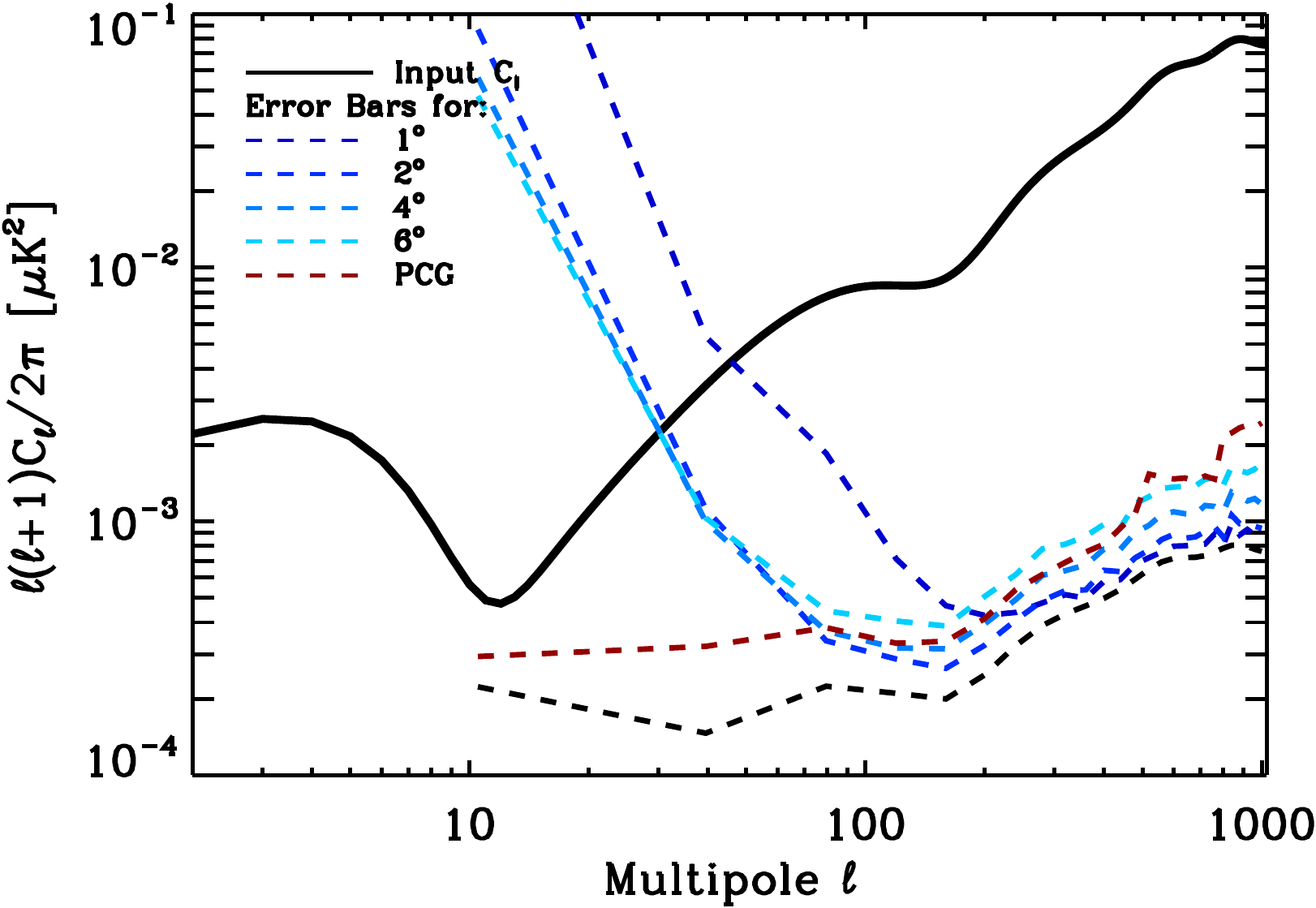}
	\caption{Statistical uncertainties on the reconstructed $B$-mode for the case of an array of telescopes. The solid-black curve stands for the input angular power spectrum with a tensor-to-scalar ratio equal to $r=0.1$. The dashed-black curve is for the binned, mode-counting uncertainties used as a benchmark. The error bars obtained by using the pixel-based, optimized sky apodizations (called PCG in the figure legend) is represented by the dashed-red curve. The different dashed-blue curves correspond to the error bars obtained by using analytic sky apodizations with an apodization length ranging from 1 degree to 6 degrees.}
	\label{fig:clhalfsky}
\end{center}
\end{figure}

An example of the uncertainties for $r=0.1$ and using the pixel-based, optimized sky apodizations (the dashed-red curve) or the analytic ones for different values of the apodization length (the dashed-blue curves) is shown in Fig. \ref{fig:clhalfsky}. The solid-black curve stands for the input angular power spectrum and the dashed-black curve stands for the binned, mode-counting computation of the error bars. This first shows that the pixel-based, optimized sky apodizations perform the best at the largest angular scales. This is systematically so for the two first bins. For the third bin, the pixel-based, sky apodizations and the analytic ones perform the same for $r>0.1$, while for $r\leq0.1$, the analytic sky apodizations with an apodization length of 4 degrees lead to a smaller error bar than the pixel-based, optimized sky apodization. However at smaller scales, $\ell\gtrsim100$, the smallest error bars are systematically obtained by using an analytic sky apodizations with an apodization length of 2 degrees for intermediate scales, $100\lesssim\ell\lesssim300$, and an apodization length of 1 degree for small scales, $\ell\gtrsim300$. We found this to be independant of the value of $r$ (at least for the grid of values considered here). \\

We note that the  apparent failure  of the pixel-based, optimized sky apodizations here may be rather due to practical difficulties in computing such apodizations sufficiently accurately, 
rather than an indication of some fundamental problems. Indeed, we have found that for the noise levels the iterative solver used to compute the apodizations converges
extremely slowly (as also observed in Ref. \cite{smith_zaldarriaga_2007}) potentially preventing us in practice from achieving sufficient precision.

%An important remark is in order here. One cannot conclude at this stage that the pixel-based, optimized sky apodizations will systematically fail in providing the smallest uncertainties at multipoles greater than $\sim100$. The iterative process of our PCG algorithm is stopped once a precision of $\epsilon<10^{-10}$ is reached\footnote{This precision at a step $n$ in the iteration is defined as $\epsilon=\left|\mathbf{A}\mathbf{W}_n-\mathbf{B}\right|$ with $\mathbf{A}$ the matrix of the linear system, $\mathbf{B}$ the vector in the right-hand-side and $\mathbf{W}_n$ the solution at the step $n$. We use the euclidean norm $L_2$.}. One then cannot exclude that smaller error bars would be achieved by computing these optimized sky apodization more accurately. However, doing this with a noise level of 1$\mu$K-arcminute leads to an unreasonnable amount of computational time (as already stated in Ref. \cite{smith_zaldarriaga_2007}).

\subsubsection{Summary on the power spectrum uncertainties} 
\begin{figure}
\begin{center}
	\includegraphics[scale=0.5]{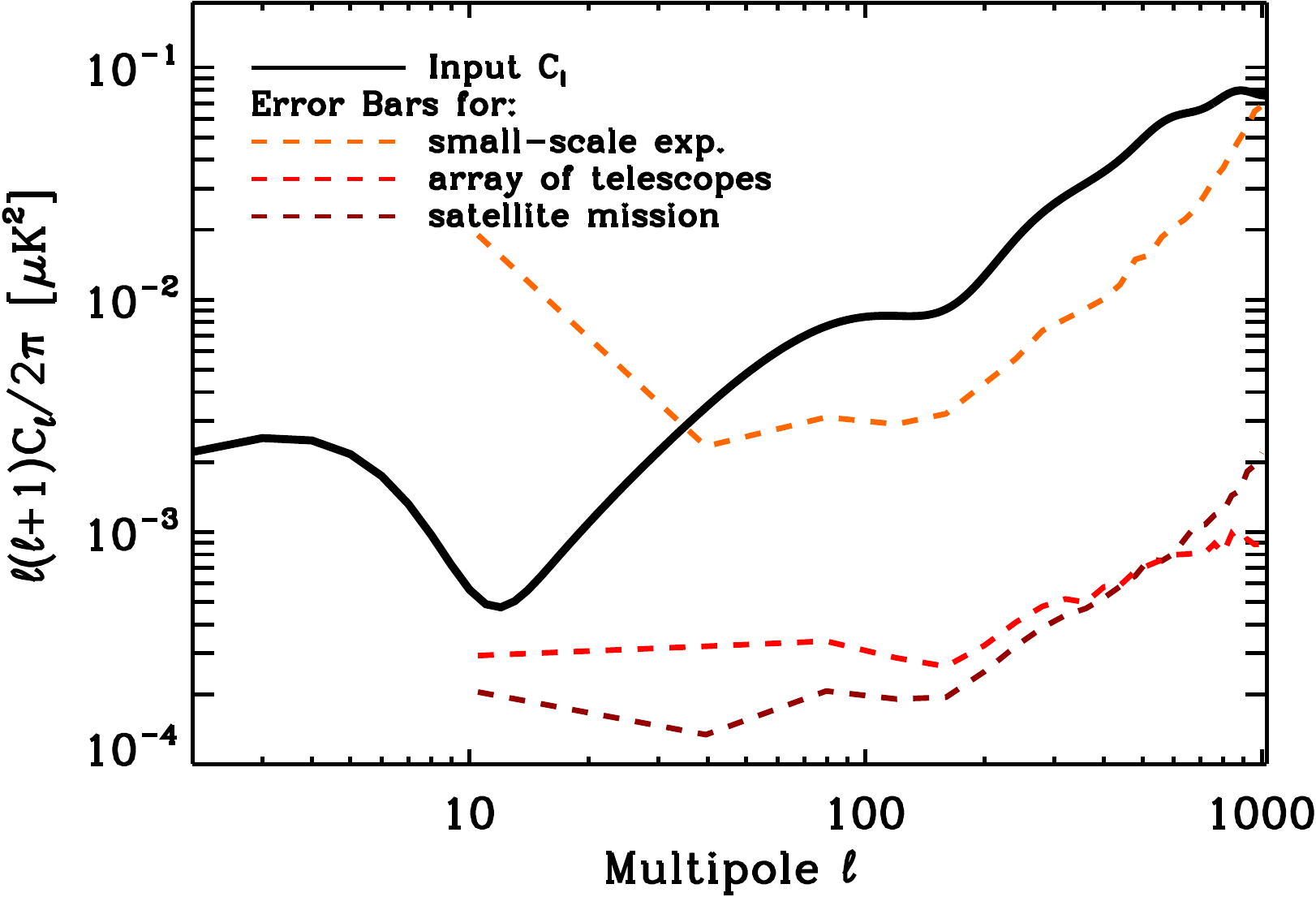}
	\caption{Statistical uncertainties on the reconstructed $B$-mode's angular power spectra with the pure pseudospectrum estimator. The orange, red and burgundy lines stand for the small-scale experiment, an array of telescopes and a satellite mission respectively. The black line corresponds to the input angular power spectrum for $r=0.1$.}
	\label{fig:cellallexp}
\end{center}
\end{figure}

As a summary, the smallest statistical uncertainties obtained for $r=0.1$ are shown in Fig. \ref{fig:cellallexp} in which the orange, red and burgundy curves stand for the small-scale experiment, an array of telescopes and a satellite mission respectively. For each experimental setup, we show the smallest error bars, which are attained for each bandpower. For the cases of a small-scale experiment and a satellite mission, this is obtained by using the pixel-based, optimized sky apodizations throughout the entire range of angular scales. For the case of an array of telescopes, the pixel-based, optimized sky apodizations are used for multipoles lower than 100, while analytic sky apodizations with an apodization of 2 degrees and 1 degree are used in the range $100\leq\ell\leq300$ and in the range $\ell>300$, respectively. \\

As expected, the higher uncertainties are the ones from a small-scale experiment due to the tiny fraction of the sky it covers, and the relatively high level of instrumental noise. We provide the uncertainties for the first bandpower, $2\leq\ell<20$, for completeness. These scales are nonetheless unaccessible starting from a map covering 1\% of the sky due to the high uncertainties, as already stated in Sec. \ref{sec:cap}.

For angular scales going from $\ell=2$ to $\ell\sim600$, the smallest error bars corresponds to a satellite mission. This is because at these angular scales, the uncertainties are dominated by sampling variance and a satellite mission, as compared to an array of telescopes, benefits from its larger sky fraction. For multipoles smaller than $\sim100$, the error bars from a satellite mission are roughly 1.5 to 2 times smaller than the error bars obtained from an array of telescopes, which is in line with the fact that the sky fraction for a satellite mission is 2 times higher than the sky fraction observed by an array of telescopes, thus reducing the error bars by a factor of $\sim\sqrt{2}$ as compared to the error bars from an array of telescope. 

Nevertheless, at small scales, $\ell>600$, smaller error bars are obtained from an observation by an array of telescopes. This is because in that regime, the uncertainties for the case of a satellite mission are dominated by the noise term, $N_\ell/B^2_\ell$. Since the noise for a satellite mission is four times higher (in power spectrum) than the noise for an array of telescopes, and the beam is more than two times higher, this increase of the variance largely overcome the decrease due to a larger sky coverage. This quantitatively explains why at those small angular scales, the lowest error bars on the $B$-mode reconstruction are obtained from an array of telescope. (One can even notice that for the range of angular scales considered here, the uncertainties obtained for an array of telescopes in sampling variance dominated.)

\subsection{Signal-to-noise ratio on $r$}
The computation of the signal-to-noise ratio on the tensor-to-scalar ratio is done by using the same Fisher matrix formalism as employed in the previous section, Eq. (\ref{eq:fisher}). For each experimental configurations and each value of $r$, we select the smallest error bars we obtained {\it bin per bin}. This means that for the specific case of an array of telescope, the estimation of the $B$-mode angular power spectra is done by mixing different kind of sky apodizations\footnote{We mention that mixing sky apodizations could lead to practical difficulties for computing the correlations between different bandpowers, though this remains conceptually similar to a case without mixing different types of sky apodizations.}. 

We use the same bandpowers as in Sec. \ref{sec:cap} and now add the largest angular scales, from $\ell=2$ to $\ell=20$ gathered in one single bandpower, in the summation in Eq. (\ref{eq:fisher}). 
Adding these scales is relevant for the case of an array of telescopes, and the case of a satellite mission. We will first add this bandpower at the largest scales for the three experimental setups, Sec. \ref{sssec:numres}. We will subsequently study its impact on the measurement of $r$, Sec. \ref{sec:firstbin}. (Note that we use the binned covariance for both the modecounting and the pure pseudospectrum reconstruction of $C^B_\ell$.)

\subsubsection{Numerical results}
\label{sssec:numres}
\begin{figure*}
\begin{center}
	\includegraphics[scale=0.375]{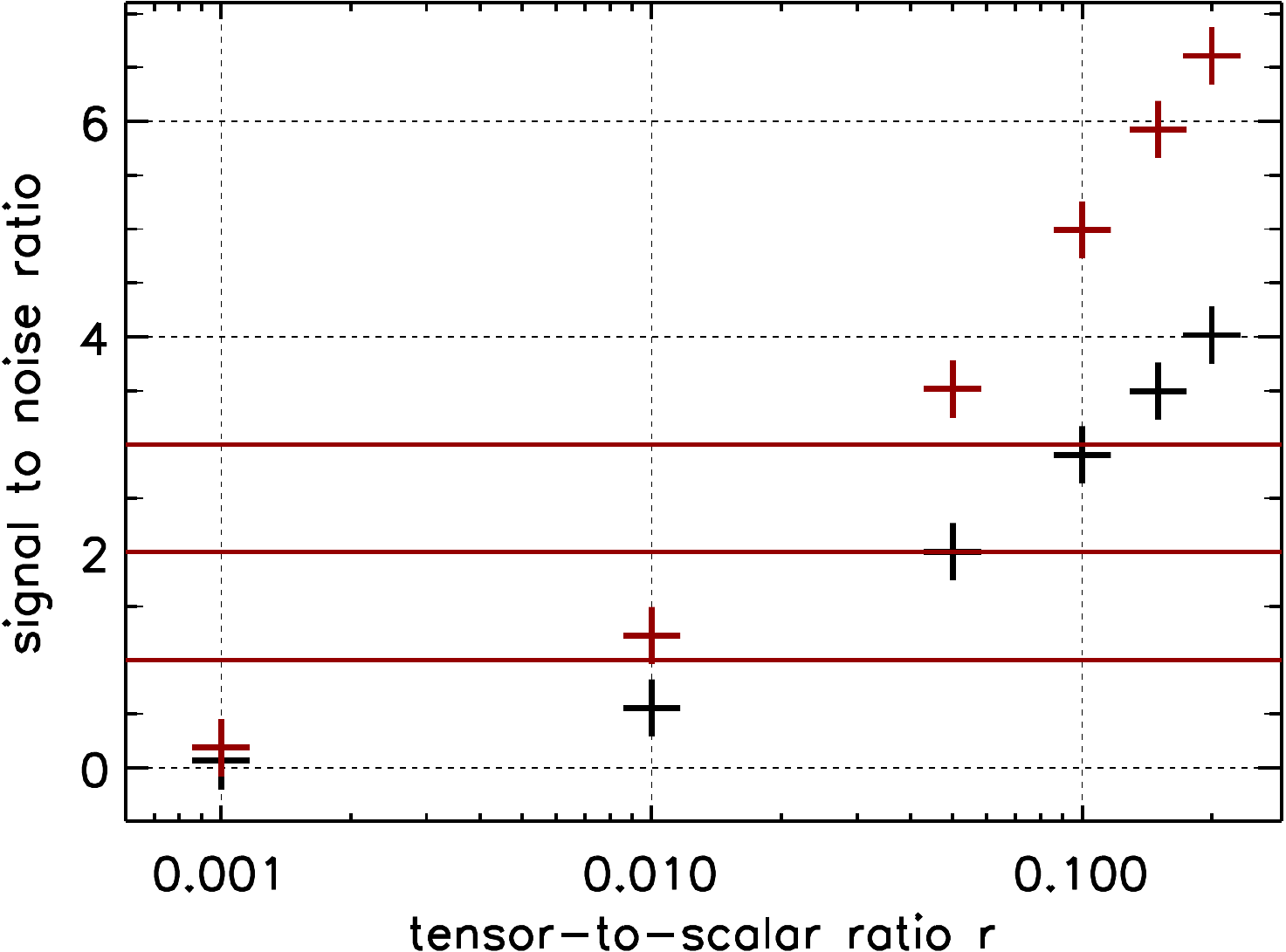} \includegraphics[scale=0.375]{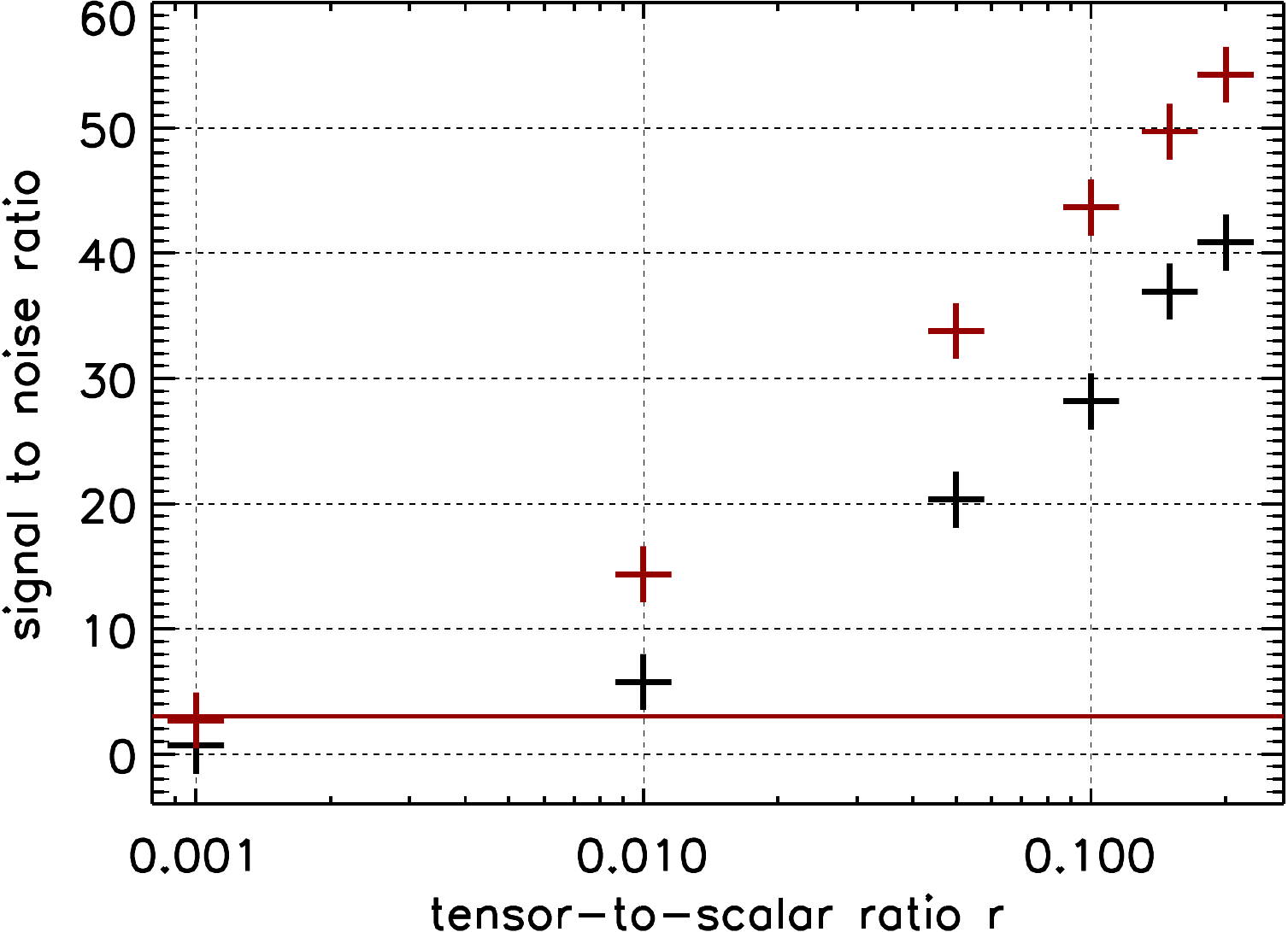} \includegraphics[scale=0.375]{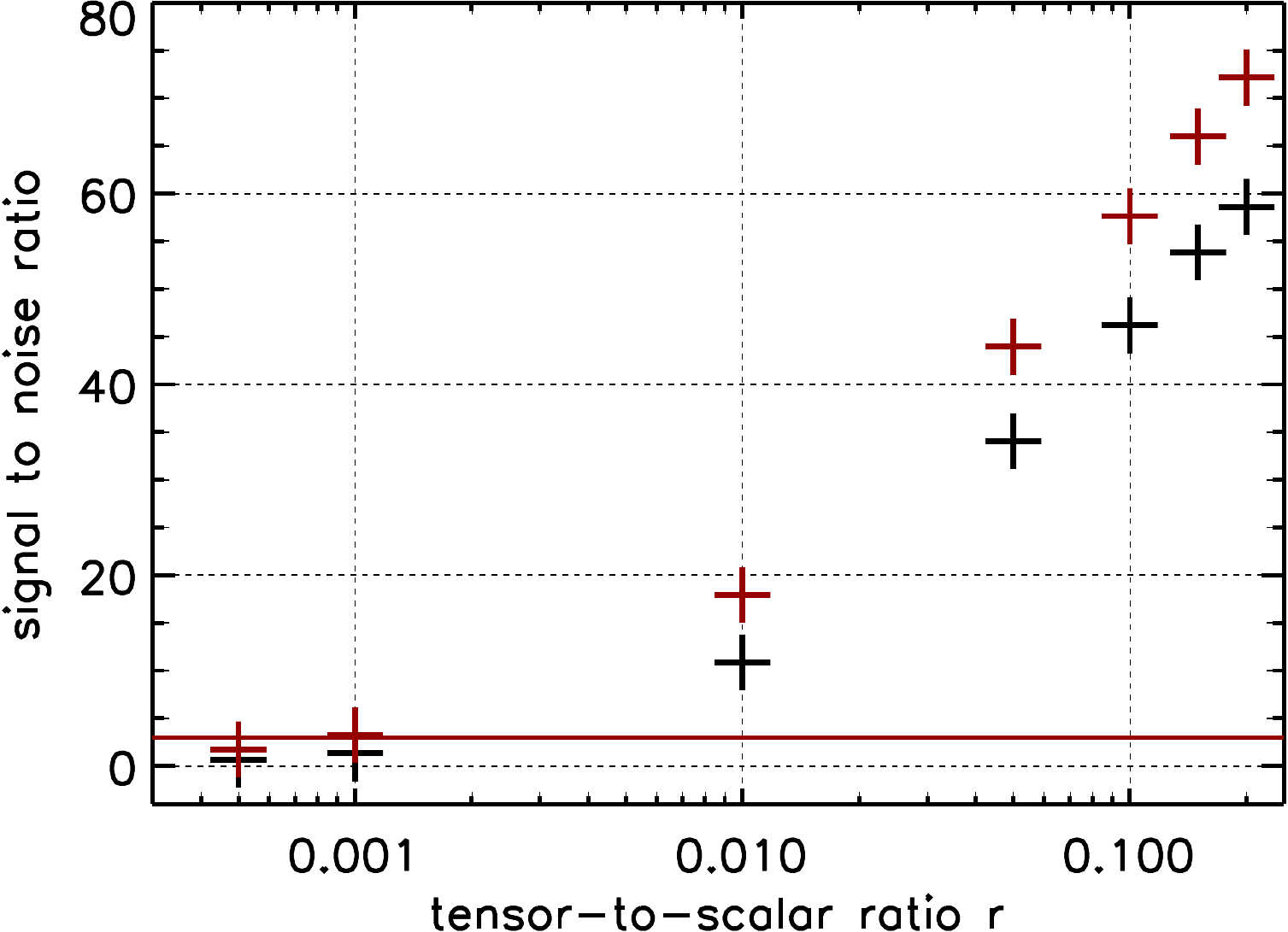}
	\caption{Signal-to-noise ratio for the detection of $r$ using pure pseudospectrum reconstruction for a potential small-scale experiment ($f_\mathrm{sky}=1\%$ and inhomogeneous noise at an average level of 5.75$\mu$K-arcminute) in the left panel, an array of telescope ($f_\mathrm{sky}=36\%$ and homogeneous noise at 1$\mu$K-arcminute) in the middle panel, and, a satellite mission ($f_\mathrm{sky}=71\%$ and homogeneous noise at 2.2$\mu$K-arcminute) in the right panel. Red crosses assume the mode-counting expression for the error bars on the reconstructed $C^{B}_\ell$ and black crosses assume the pure pseudospectrum error estimated using Monte-Carlo simulations.}
	\label{fig:snrallexp}
\end{center}
\end{figure*}

Our results on the signal-to-noise ratio for $r$ ranging from $0.001$ to $0.2$ are shown in Fig. \ref{fig:snrallexp} (note that for the specific case of a satellite mission the value $r=5\times10^{-4}$ has been added in order to fall below the $3\sigma$ limit). The red and black crosses correspond to a covariance matrix computed using the mode-counting expression for error bars on $C^{B}_\ell$, and the pure pseudospectrum error bars, respectively. The solid red line is the $3\sigma$ limit. The left panel corresponds to a small-scale experiment covering $\sim$1\% of the celestial sphere with a highly inhomogeneous noise distribution. The middle panel corresponds to an array of telescopes covering $\sim$36\% of the sky with a low level of (homogeneous) noise. Finally, the right panel is for a satellite mission covering $\sim$71\% of the sky with a low level of homogeneous noise. \\

For the case of a small-scale experiment, the signal-to-noise ratio on $r$ ranges from 0.06 for $r=0.001$ to 4 for $r=0.2$ assuming a pure pseudo-$C_\ell$ reconstruction of the $B$-mode power spectrum (meaning that a "measurement" of $r=0.001$ would be consistent with $r=0$). 
This has to be compared to what would be inferred from the idealized mode-counting evaluation of the uncertainties, for which the signal-to-noise ratio varies from 0.25 for $r=0.001$ to 6.7 for $r=0.2$. For $r=0.001,~0.01$ and $0.1$, the (S/N)$_r$ derived from a mode-counting estimation of the uncertainties on the $B$-mode is overestimated by a factor $4.2$, $2.3$ and $1.8$, resp., as compared to the (S/N)$_r$ obtained from a pure pseudo-$C_\ell$ reconstruction of the angular power spectrum.

For the case of an array of telescopes and assuming the pure pseudosepctrum estimation of the $B$-mode, the signal-to-noise ratio varies from 0.67 to 41 with $r$ varying from 0.001 to 0.2. Values of $r=0.01$ and $r=0.1$ would be measured with a statistical significance of 5.75$\sigma$ and $28.16\sigma$, respectively. Using instead the mode-counting estimation of the uncertainties on $C^{B}_\ell$, the (S/N)$_r$ varies from 3 to 54 for $r$ ranging from 0.001 to 0.2. For the three selected values of $r=0.001,~0.01$ and 0.1, the signal-to-noise ratio obtained from the mode-counting approach is respectively overestimated by a factor 4.5, 2.5 and 1.3, as compared to the realistic (S/N)$_r$ derived from the pure pseudo-$C_\ell$ estimation of $C^{B}_\ell$.

For the case of a satellite mission, the signal-to-noise ratio varies from 0.66 for $r=0.0005$ to 59 for $r=0.2$, and assuming the pure pseudospectrum estimation of $C^{B}_\ell$. The values of $r=0.001$, $r=0.01$ and $r=0.1$ would be detected with a statistical significance of 1.34, 10.84 and 46.19, respectively. If one instead makes use of the mode-counting estimation of the error bars on the $B$-mode reconstruction, the (S/N)$_r$ varies from 2 to 72 for values of the tensor-to-scalar ratio ranging from 0.0005 to 0.2. For $r=0.001,~0.01$ and $0.1$, the mode-counting evaluation overestimates the signal-to-noise ratio, as compared to the pure pseudo-$C_\ell$ reconstruction, by a factor 2.2, 1.59 and 1.22. \\

From a qualitative viewpoint, the signal-to-noise ratios computed in the framework of the mode-counting expression are always higher compared to the signal-to-noise ratios assuming the pure pseudospectrum reconstruction of $B$-mode. (This is obviously expected from the fact that the mode-counting approach is an idealized and {\it underestimated} computation of the uncertainties.) We observe that the overestimation using the mode-counting expression (as compared to the more realistic pure pseudospectrum reconstruction of $C^{B}_\ell$) is less marked for higher values of $r$. This behavior is common to the three experimental configurations here-considered, though there are differences from a quantitative viewpoint. The reason is that for low values of $r$, most of the information comes from the largest scales, which is precisely at those large scales that the underestimation of the $B$-mode reconstruction using the mode-counting formul\ae~is more marked.

We also stress that in the case of mode-counting approach, the leakages are ignored. 
On the contrary, the pure pseudospectrum approach consistently includes them but correct them in the analysis.
This explains why the mode counting approach overestimate the signal-to-noise ratio on $r$.

\subsubsection{Relative importance of the reionization peak}
\label{sec:firstbin}
CMB observations covering a large fraction of the sky are automatically contaminated by various astrophysical foregrounds with complex physics involved among which the emission from our galaxy is the strongest. Masks are used to remove from the analysis the portion of sky with the highest foreground level, but the foreground emission is present on the entire celestial sphere. Usually techniques - such as parametric component separation \cite{stompor_etal_2009} used to determinate the spectral parameters or template fitting method, which deprojects the template of the foreground from the map \cite{jaffe_etal_2004,katayama_etal_2011} - are used to minimize the impact of the foreground. The residual level of foreground contaminants depends on the technique actually chosen. However, the power spectrum of the galactic dust, polarized emission (which is the major contaminant of CMB measurements at frequencies above $\sim100$GHz) behaves as $\ell^{-2.4}$, to be compared to $\ell^{-2}$ for the CMB $B$-mode angular power spectrum at scales above a degree \cite{planck_dust}. The impact of such a galactic foreground is therefore expected to be more pronounced at the largest angular scales.

Here, we considered the worst case scenario where the foreground contamination could not be removed at all on the largest scale, meaning that the information from the reionization peak is no more taken into account in the computation of the signal-to-noise ratio. In practice, we discard the first bin ($2\leq \ell<20$) from the analysis, which necessarily lowers the signal-to-noise ratio on $r$. We define this relative decrease as:
\begin{equation}
	\delta=\left|\frac{\mathrm{(S/N)}_r^{(\ell>20)}-\mathrm{(S/N)}_r}{\mathrm{(S/N)}_r}\right|,
\end{equation}
with (S/N)$_r$ the signal-to-noise on $r$ accounting for {\it all} the angular scales, and (S/N)$_r^{(\ell>20)}$ the signal-to-noise ratio obtained by {\it discarding} the first bandpower. This relative decrease can alternatively be interpreted as the relative contribution from the first bin to the signal-to-noise on $r$ since:
\begin{equation}
	\delta=\left|\frac{\mathrm{(S/N)}_r^{(\ell<20)}}{\mathrm{(S/N)}_r}\right|,
\end{equation}
with (S/N)$_r^{(\ell<20)}$ the signal-to-noise ratio on $r$ that would be obtained by using the first bandpower {\it only}. \\

\begin{figure*}
\begin{center}
	\includegraphics[scale=0.35]{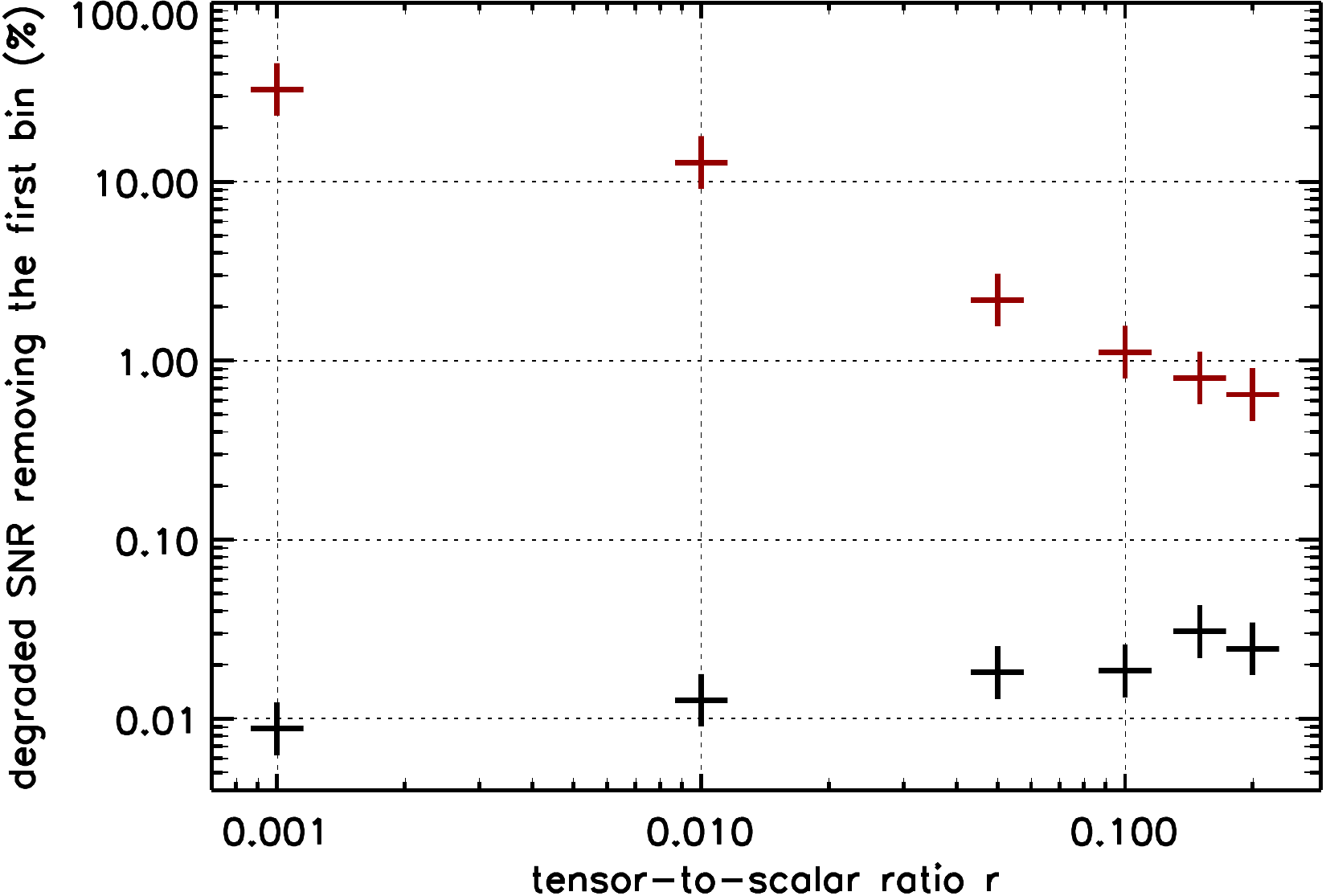}\includegraphics[scale=0.35]{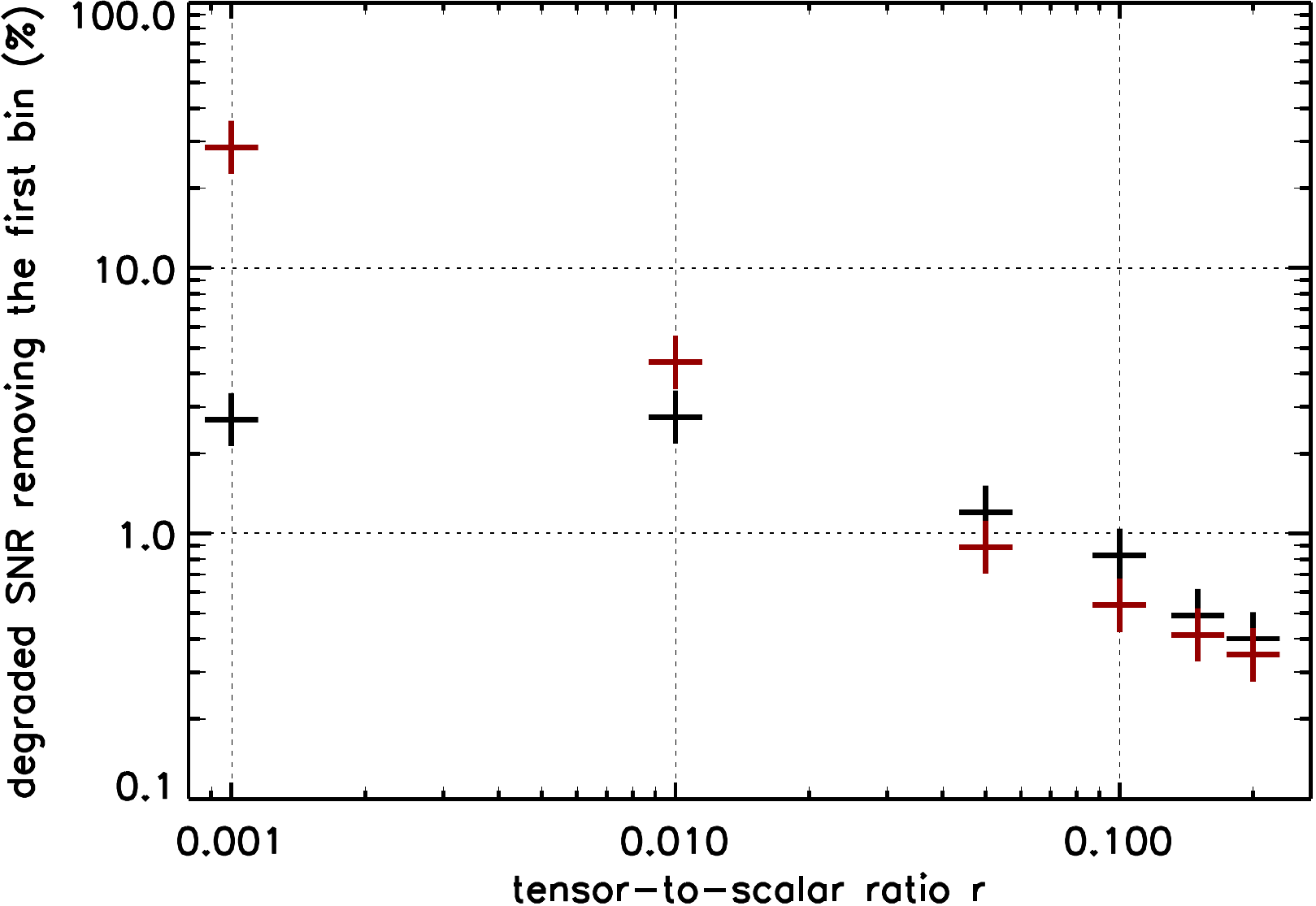}\includegraphics[scale=0.35]{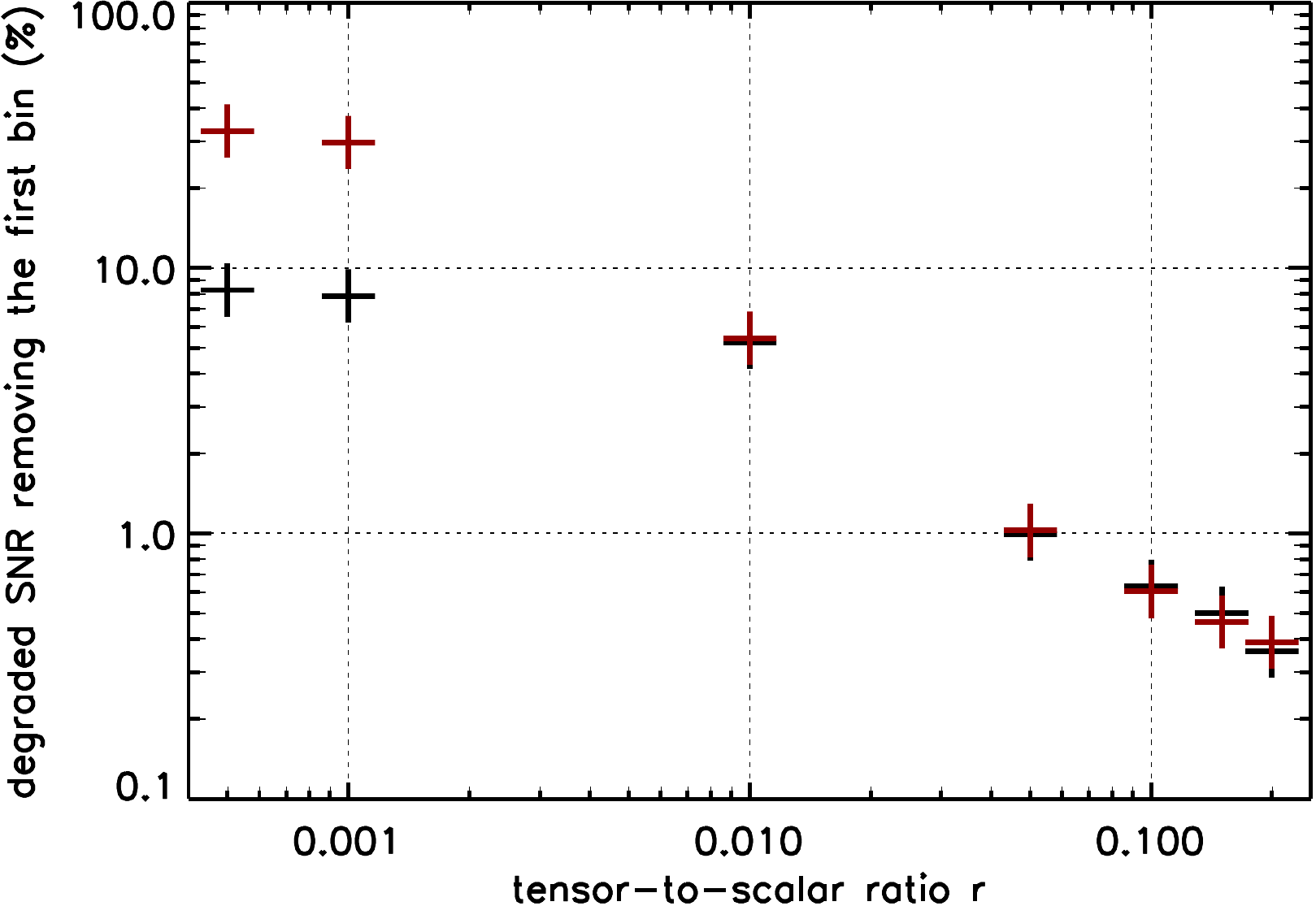}
	\caption{Relative decrease of the signal-to-noise ratio (in percent) if the information from the reionization peak (i.e. the first bin) is discarded from the analysis. The left, middle and right panels are for the small-scale experiment, an array of telescopes and a satellite mission, respectively.}
	\label{fig:degradesnrallexp}
\end{center}
\end{figure*}

This relative decrease of (S/N)$_r$ is shown in Fig. \ref{fig:degradesnrallexp}. The red crosses correspond to the mode-counting estimation of the error bars on the reconstruction of $C^{B}_\ell$ while the black crosses correspond to the error bars from a pure pseudospectrum estimation of $C^{B}_\ell$. The left, middle and right panels respectively stand for the case of a small-scale experiment, an array of telescopes and a satellite mission. 

The case of a small-scale experiment is poorly affected by the removal of the first bin using the pure peudospectrum reconstruction of $C^{B}_\ell$, the relative decrease being systematically smaller than 0.1\%. This is because in such a case the signal-to-noise ratio for the first bandpower, $C^{B}_{b=1}/\sqrt{\mathbf{\Sigma}_{b=1,b=1}}$, is much smaller than unity for all the values of $r$ considered here. This bandpower therefore does not bring any significant amount of informations on $r$. This drastically differs if one uses the mode-counting evaluation for which $\delta$ varies from 0.7\% for $r=0.2$ to 32\% for $r=0.001$. This is because in this case, the signal-to-noise ratio in the first bandpower, $C^{B}_{\ell}/\sqrt{\mathbf{\Sigma}_{\ell,\ell}}$ with $2\leq\ell<20$, becomes greater than unity though the sky coverage is only of $\sim1\%$. The fact that the relative decrease is more pronounced for small values of $r$ is understood as follows. For lower values of $r$, the recombination bump at the degree scale, falls below the lensing part of the $B$-mode while the reionization bump in the first bandpower remains above the lensing signal. As a consequence, the reionization peak carries more information, relative to the informations carried by the recombination peak, for lower values of the tensor-to-scalar ratio.

For the case of an array of telescopes, the relative decrease ranges from 0.4\% for $r=0.2$ to roughly 3\% for $r=0.001$. We note that this relative decrease is now roughly constant from $r=0.001$ to $r=0.01$ and then decreases for higher values of the tensor-to-scalar ratio. This behavior of $\delta$ is explained by the very same reason explaining why $\delta$ decreases for higher values of $r$ if one makes use of the mode-counting estimation of the uncertainties on the estimated $C^{B}_\ell$, and because for an array of telescopes, the angular power spectrum in the first bandpower can now be measured with a signal-to-noise ratio greater than unity. We note that the relative decrease using the mode-counting behaves the same as in the case of a small-scale experiment (with minor quantitative differences at high values of $r$).

For the case of a satellite mission, the relative decrease varies from 0.35\% for $r=0.2$ to 9\% for $r=0.0005$, with $\delta=8\%$ for $r=0.001$. This relative decrease now monotonically increases with lower values of $r$ (though our results suggest that a plateau is reached for $r<0.001$). This behavior is explained by the same reason explaining the decrease of $\delta$ for higher values of $r$ in the case of an array of telescopes. We also note that $\delta$ obtained from the mode-counting expression behaves the same as in the case of a small-scale experiment and an array of telescopes. \\

As is clear from Fig. \ref{fig:degradesnrallexp}, the shape of $\delta$ as derived using the mode-counting expression, is qualitatively the same for the three experimental configurations, though sky fractions and shapes of the masks drastically change. This is because the impact of the limited sky fraction is simply modelled as an overall renormalization of the error bars, equally applied at all angular scales (see Eq. (\ref{eq:mcount})). Neglecting the noise contribution to the error bars on the $B$-mode reconstruction (which is a relatively fair assumption here), it is easy to figure out that this overall $1/\sqrt{f_\mathrm{sky}}$ does not enter in the final expression of $\delta$. (We note that minor differences are however expected because of the different noise level and beamwidth.) At low values of $r$, the relative decrease is much less marked in the context of the pure pseudospectrum reconstruction of $C^{B}_\ell$, with respect to the mode-counting expression. This is because the different leakages have stronger impacts at large scales (in term of increase of the error bars on the estimated $C^{B}_\ell$), thus reducing the relative contribution of the first bandpower to the constraint that can be set on $r$. The impact of leakages in terms of error bars on $C^{B}_\ell$ at large angular scales increases with smaller $f_\mathrm{sky}$, which therefore reduces the relative contribution of the first bandpower to the constraints on $r$. This is clearly seen in Fig. \ref{fig:degradesnrallexp} where $\delta$ is more important from the case of a small-experiment, to the case of an array of telescopes, to the case of a satellite mission.

\subsubsection{Performances on $r$ detection}
As a result, at a given $r$ and considering all the angular scales from 2 to 1020, the value of the signal-to-noise ratio is the highest in the case of a satellite mission. As an example, a tensor-to-scalar ratio $r = 0.1$ would be detected at a statistical significance of about $46\sigma$.  In the case of an array of telescope, the value of the signal-to-noise ratio remains high for a large range of values of the tensor-to-scalar ratio, showing a detection of 28$\sigma$ at $r\sim0.1$. Finally, a small scale experiment would set mild constraints on low values of the tensor-to-scalar ratio, reaching 3$\sigma$ at $r\sim0.1$. \\

In the frame of the primordial $B$-mode detection prospects, the minimal value of the tensor-to-scalar ratio $r$ that could be detected regarding the experimental setups is a relevant result. The table \ref{tab:snr} summarizes in this perspective the aforementioned results, considering a measurement of $r$ with at least a 3$\sigma$ statistical significance as a threshold. The minimal accessible value of $r$ is shown with respect to the experimental setups and the estimation of the $B$-mode variance over all the range of multipoles (referred to as {\it case A} in the table). As explained above, the mode counting estimation of the variance overestimates the forecasts made on the minimal accessible $r$ as compared to the realistic $B$-mode estimation. In the case of a potential satellite mission for instance, the lowest accessible $r$ we could realistically expect is 2.88 greater than the one estimated using the mode-counting estimation. 
These results therefore highlight the inaccuracy that an approximative estimation of the $B$-mode induces on the performed forecasts of the detectable $r$ values. Thus considering the realistic forecasts performed thanks to the pure estimation of the $B$-mode power spectrum, we conclude that a satellite mission would give access to the largest range of $r$, with a minimal $r$ value of $2.6\times10^{-3}$. A typical small scale experiment is indeed expected to reach only $r$ higher than $0.1$ at 3$\sigma$ (note that $r=0.05$ is detectable at $2\sigma$). Between these two cases lies the one of an array of telescopes, which warrants a detection of the tensor-to-scalar ratio if it is higher than $5.1\times10^{-3}$. As a result, each studied experiments widens the accessible range of the tensor-to-scalar ratio $r$. In terms of minimal detectable vlaue of $r$, one gains about a factor 20 from small-scale experiments to an array of telescopes, and about a factor 2 between the latter and a satellite mission.  \\

\begin{table}
\begin{center}
\begin{tabular}{p{2.4cm}m{1.9cm}m{1.9cm}m{1.9cm}} \hline\hline
          &  Small-scale exp. & Telescopes array  & Satellite mission   \\
     \hline
     Mode-counting: &  &  & \\
     ~~~Case A & $r \gtrsim 0.038$ & $r \gtrsim 0.0011$ & $r \gtrsim 0.0009$ \\
     ~~~Case B & $r \gtrsim 0.04$ & $r \gtrsim 0.0016$ & $r \gtrsim 0.0013$ \\ \hline
     %&&& \\
     Pure pseudo-$C_\ell$ : & & & \\
     ~~~Case A  & $r \gtrsim 0.11$     & $r \gtrsim 0.0051$ & $r \gtrsim 0.0026$ \\
     ~~~Case B  & $r \gtrsim 0.11$     & $r \gtrsim 0.0053$ & $r \gtrsim 0.0028$ \\
     \hline\hline
\end{tabular}
     \caption{The minimal accessible value $r$ with {\it at least} a $3\sigma$ statistical significance, regarding the experimental setups and the estimation of the variance on the $B$-mode reconstruction. The {\it case A} means that all the bins are used. The {\it case B} means that the information from the reionization peak (i.e. the first bin from $\ell=2$ to $\ell=20$) has been removed. This is obtained by linearly interpolating the computed (S/N)$_r$ on our grid of values of $r$.} 
     \label{tab:snr}
\end{center}
\end{table}

Furthermore, the table \ref{tab:snr} also displays the minimum accessible $r$ obtained without the information from the first bin ($\ell < 20$) of the $B$-mode power spectrum (referred to as the {\it case B}). This lack of information obviously leads to a smallest range of accessible $r$ than in the {\it case A} for a naive estimation of the $B$-mode variances, as explained in the previous subsection. In particular, while the accessible $r$ range is little affected by removing the first bin in the case of a small scale experiment, the minimal accessible $r$ is $\sim 1.44$ ($\sim 1.04$ resp.) greater for a satellite mission (an array of telescopes resp.) as the largest angular scales are relevant for these experimental setups. We note here that contrary to what one might expect, a large scale experiment would still succeed in detecting $r$ of at least $10^{-3}$. For $\ell$ between 20 and 90, the amplitude of the primordial signal is roughly 10\% of the lensing signal while the total (mode-counting estimated) error budget varies from few percents to 10\% of the lensing signal. Summing over the multipoles range thus enables a detection of $r\sim10^{-3}$ with a $3\sigma$ statistical significance.  

Nonetheless, in this {\it case B}, the orders of magnitude of the realistic forecasts remain unchanged if the $B$-mode power spectrum is reconstructed from the pure pseudo-$C_\ell$ approach. The values of $r$ that could be detected at $3\sigma$ increase by a factor of less than one percent for a small-scale experiment, a factor of few percents for an array of telescopes, and, a factor of ten percents for a satellite mission. (This obvisouly reflects the values of $\delta$ found in the previous section, Sec. \ref{sec:firstbin}.) This means that the pure pseudo-$C_\ell$ estimation of the reionization peak of the $B$-mode mildly constraints the tensor-to-scalar ratio. To take full advantages of the range $2\leq\ell\leq20$ (so as to lower the minimal detectable value of $r$ and to enlarge the lever arm to constraint e.g. the spectral index), one should probably rely on more optimal techniques for reconstructing $C^{B}_\ell$ at those largest angular scales\footnote{We note that the reconstruction of $C^{B}_\ell$ at large scales is also plagued by other sources of uncertainties such as the level of residual foregrounds and/or the impact of filtering of the maps.}. 

%%%%%%%%%%%%%%%%%%%%%%%%%%%%%%%%%%%%%%%%%%%%%%%%%
\section{Conclusion and discussion}
\label{sec:conclu}
We have investigated the detection of the tensor-to-scalar ratio, $r$, from forthcoming and potential future measurements of the CMB polarized anisotropies. We considered the $B$-mode as the main source of information on $r$ and assumed the pure pseudospectrum reconstruction of its angular power spectra from the maps of Stokes $Q$ and Stokes $U$, previously shown to be a method of choice for analyzing coming data sets. We focused on realistic statistical uncertainties (i.e. sampling and noise variance) as incurred by such a numerical method, and we purposefully did not consider the potential gain thanks to delensing, nor the loss due to polarized foreground contamination and instrumental systematics. 

We emphasize that in this paper we consider only the $E$-to-$B$ leakage due to a cut sky. In CMB practice there are numerous other potential sources of such leakages. For instance, they can arise from instrument limitations, such as beam mismatch~\cite{shimon_etal_2008} or polarimeter orientation uncertainty ~\cite{keating_etal_2012}, or be generated by data processing, say, via time-domain filtering ~\cite{bicep2_2014,pbr3}. Such leakages would also have an effect on estimated $B$-mode power spectrum. The effect will in general depend on a specific method used for the estimation but also on the detailed nature of the leakage itself,
and thus would have to be studied cased by case. In many situations, such leakages could be corrected for already on the map-making stage, leaving therefore
 the cut-sky as the only fundamental source of the leakage to contend with on the power spectrum estimation level as assumed in this work. 
 
In contrast, we include the effects of the gravitational lensing, i.e., of the "cosmological $E$-to-$B$ leakage", in the total uncertainty budget, in spite of the fact that map-making-level, delensing procedures, which could correct for part of this effect have been proposed~\cite{delens}. The improvements on the detection of $r$ those methods could give depend on the noise level and the resolution of the experiment, and, on the potential use of external datasets (if delensing cannot be done internally). By including the lensing-induced $B$-mode, we adopt a more conservative viewpoint as far as forecasts on the tensor-to-scalar ratio are concerned.\\

%{\bf We also mention that the pure pseudospectrum estimator presently used solves only for the $E$-to-$B$ leakages due to cut sky effects. We do not consider here any additional leakages due to instrumental effects (e.g. beam mismatch or filtering of the time streams) which could require amendments of the pure pseudospectrum formalism.} \\

In this framework, we first consider the case of small-scale (either ground-based or balloon-borne) experiments in an idealized way, assuming the observed sky patch is azimuthally symmetric. We consider four values of $r=0.07,~0.1,~0.15$ and $0.2$, and let the sky fraction to vary from 0.5\% to 10\% (with a noise level of $5.75\mu$K-arcminute at $f_\mathrm{sky}=1\%$. We compare the signal-to-noise on $r$ as obtained from the pure pseudospectrum reconstruction of the $B$-mode to the signal-to-noise ratio that would obtained assuming either the mode-counting estimation of the uncertainties on the $B$-mode, or a minimum variance, quadratic estimator. We show that the statistical significance on the detection of $r$ using the mode-counting is overestimated by a factor $\sim1.25$ as compared to the more realistic case of the pure pseudospectrum estimation. (The mode-counting also overestimate this significance by a factor $\sim1.17$ as compared to the minimum variance, quadratic estimator.) Similarly, the (S/N)$_r$ obtained from the pure pseudospectrum estimator is reduced by 1.5\%( at $r=0.2$) to 8\% (at $r=0.07$) as compared to the lossless minimum variance, quadratic estimator. 

For the case of small-scale experiment for which the reionization bump is not accessible, and in the limitation of azimuthally symmetric patches, the pure pseudospectrum approach for $B$-modes reconstruction is thus almost as accurate as the more computationally costly minimum variance, quadratic estimator (the former scaling as $\mathcal{O}(N^{3/2}_\mathrm{pix})$ and the latter as $\mathcal{O}(N^{3}_\mathrm{pix})$ if the observed sky patch is not azimuthally symmetric). As shown in Fig. 20 of \cite{grain_etal_2009}, non-azimuthal symmetry basically does not change the uncertainties on the $B$-mode reconstruction with the pure pseudospectrum estimator (except for unrealistic, highly squeezed shapes). We can thus except this conclusion to holds for more intricate shapes of the observed sky.

Our results (summarized in Tab. \ref{tab:fsky}) show that for a given sensitivity typical of forthcoming small-scale experiments, the value of the sky fraction maximizing the signal-to-noise ratio on $r$ is rather insensitive to the method adopted to compute the uncertainties on the reconstructed $C^{B}_\ell$ (either the mode-counting expression, a minimum-variance quadratic estimator or the pure pseudo-$C_\ell$ approach). We also show that the choice of the bandwidth only mildly affect this optimized sky fraction in the case of the mode-counting approach to estimate the uncertainties on the $B$-mode reconstruction (see Tab. \ref{tab:bin}). This means that using the mode-counting expression provide a rather reliable estimate of the optimized sky fraction from the viewpoint of statistical uncertainties though being underestimated. \\

Second, we consider the detection of the tensor-to-scalar ratio for three selected examples, each of them mimicking three archetypal experimental configurations. Realistic sky coverage (with intricate contours) and realistic noise distribution for the small-scale experimental setup are considered and the statistical uncertainties on the $B$-mode reconstruction are derived from the mode-counting expression first (used as a benchmark), and second, from  the pure pseudospectrum estimators using optimized sky apodizations. Our results are summarized in the table \ref{tab:snr}. For each experimental setups, it shows the minimal values of $r$ that could be measured with at least a statistical significance of $3\sigma$. One gains more than one order of magnitude for the minimal detectable value of $r$ from the small-scale experiment to an array of telescopes, and another factor 2 from an array of telescopes to a satellite mission. This conclusion stands even if the largest angular scales ($\ell \leq 20$) cannot be used in the analysis.

Let us briefly discuss the impact of those results in the context of single-field, slow-roll inflation. Our purpose here is to give a rough translation of the potential measurement of the tensor-to-scalar ratio with the pure pseudo-$C_\ell$ estimation of the $B$-mode, into a potential discrimination between small fields and large fields models of inflation. (A more detailed study of inflationary models can be found in Ref. \cite{martin_etal_2014}, though it is restricted to satellite missions and assumes a different evaluation of the error budget for the $B$-mode reconstruction.) 

The tensor-to-scalar ratio is an valuable source of information for the physics of the primordial Universe. First, it is a direct measure of the energy scale of inflation, $V^{1/4}$ with $V$ the value of the inflaton potential during inflation:
\begin{equation}
	V^{1/4}=1.06\times 10^{16}\mathrm{GeV}\left(\frac{r}{0.01}\right)^{1/4}.
\end{equation}
This means that a measured value of $r\geq0.01$ corresponds to test a physical regime in the playground of Grand Unified Theories. Second, the tensor-to-scalar ratio is directly related to the number of e-folds, $N_\mathrm{inf}$, and the excursion of the scalar field, $\Delta\phi$, from the instant when cosmological fluctuations observed in the CMB are created during inflation, to the end of inflation \cite{lyth}:
\begin{equation}
	N_\mathrm{inf}=\sqrt{\frac{8}{r}}\times \frac{\Delta\phi}{m_\mathrm{Pl}},
\end{equation}
with $m_\mathrm{Pl}=1/\sqrt{8\pi G}$ the reduced Planck mass. (We note that $N_\mathrm{inf}$ can be determined from the knowledge of the inflaton potential. We however let it free in order not to assume a too specific shape of this potential.) Single field inflationary models can be roughly classified between large fields models and small fields models, whether the excursion of the scalar field is transplanckian or subplanckian, respectively. Though the value $\Delta\phi/m_\mathrm{Pl}=1$ should not be considered as a sharp and univoquely defined frontier between small fields and large fields models\footnote{The distinction between large fields and small fields models of inflation is usually thought in the context of effective field theory and the UV completion of gravity. It is therefore natural, but not necessary, to introduce a cut-off of the order of the Planck scale.}, a precise measure of $r$ then allows for discriminating between this two classes of models. For $N_\mathrm{inf}\sim30$ and considering zero runnings of the spectral index (see \cite{huang_2015} for extensions of the Lyth bound with runnings), values of $r$ greater than $\sim0.01$ would correspond to large fields models of inflation (see also Ref. \cite{chatterjee} and references therein for examples of small fields models evading the Lyth bound).
\begin{figure}
	\includegraphics[scale=0.55]{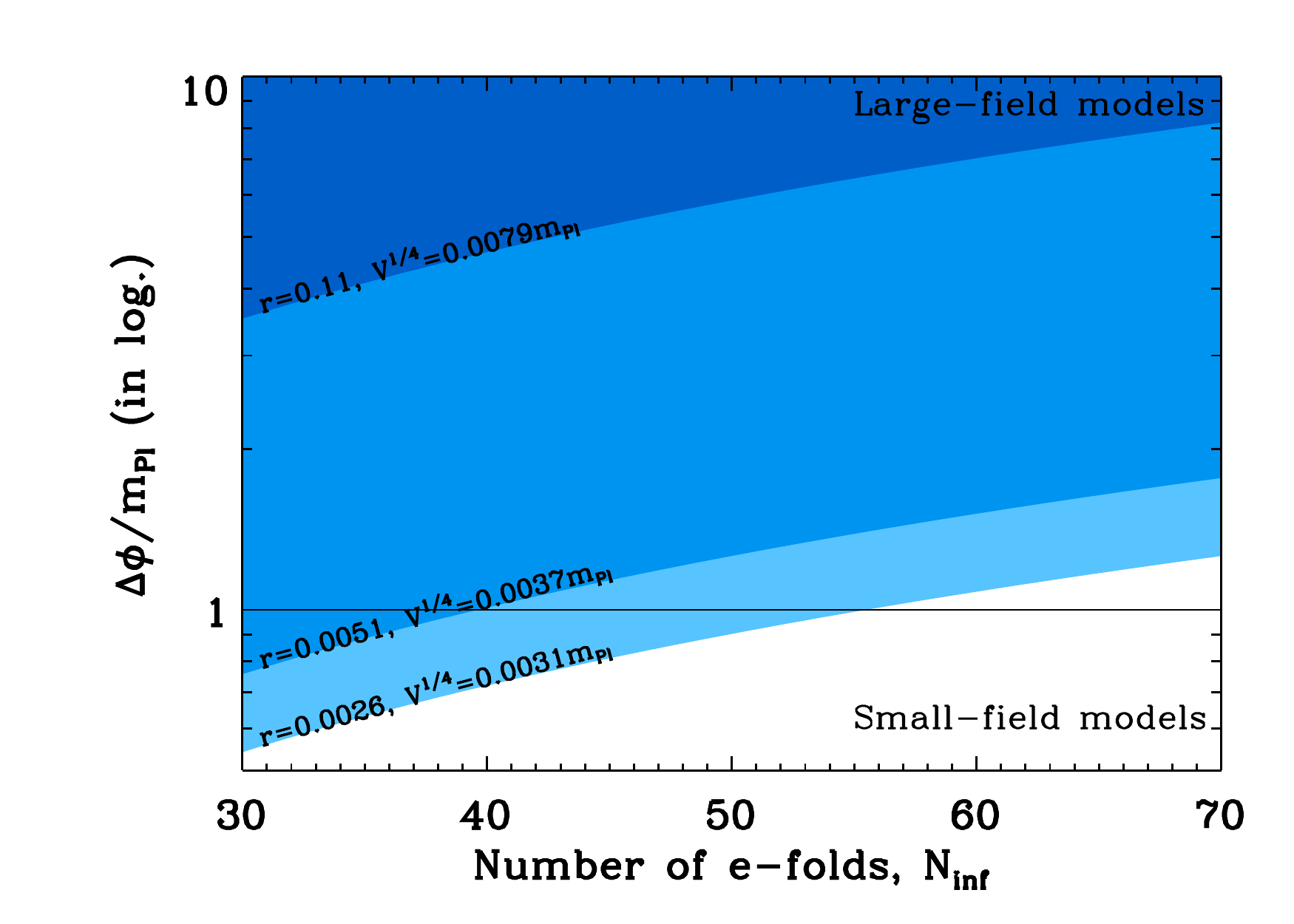}
	\caption{Values of the excursion of the scalar field, $\Delta\phi$, that could be observed with, at least, a $3\sigma$ significance, as functions of the number of e-folds during inflation. Darker blue to lighter blue respectively stands for small-scale experiment, an array of telescopes and a satellite mission. The minimal, detectable value of $r$ at $3\sigma$ allowing for such a measurement is the one derived from the pure pseudo-$C_\ell$ estimation of the $B$-mode angular power spectra.}
	\label{fig:lyth}
\end{figure}

The figure \ref{fig:lyth} shows the ranges of $\Delta\phi/m_\mathrm{Pl}$ as a function of $N_\mathrm{inf}$ accessible assuming that the tensor-to-scalar ratio has been measured with at least a $3\sigma$ statistical significance. Blue areas correspond to the accessible range for each experimental configurations (notice that the higher $\Delta\phi$, the higher $r$). The dark blue region is for the case of small-scale experiments, while the somewhat lighter blue and light blue regions corresponds to the case of the array of telescopes and of the satellite mission
respectively. The minimal detectable value of $r$ with at least $3\sigma$ is the one derived from a pure pseudospectrum reconstruction of the angular power spectra of the $B$-mode and using the entire set of angular scales (the {\it case A} of Tab. \ref{tab:snr}). This shows that a measurement of $r$ from the pure pseudo-$C_\ell$ reconstruction of the $B$-mode thanks to datasets coming from a small-scale experiment, is impossible if small fields models appear to be realized in the early Universe. Though a detection is possible in the large field models, there is still a range of such models for which the level of primordial gravity waves is still undetectable by a small-scale experiment. Small fields models are only marginally accessible from the pure pseudo-$C_\ell$ estimation of the $B$-mode using datasets from an array of telescopes, as $\Delta\phi/m_\mathrm{Pl}\leq1$ is accessible for $N_\mathrm{inf}$ smaller than $\sim38$. A detection of $r$ consistent with zero with a $3\sigma$ confidence level implies an excursion of the scalar field (in Planck units) smaller than 0.8 to 1.8 for $N_\mathrm{inf}$ varying from 30 to 70. Finally, datasets coming from a satellite mission allows for a detection of primordial gravity waves in the small fields models with the pure pseudospectrum estimation of $C^{B}_\ell$, providing that the number of e-folds is smaller than $\sim55$. On the range of e-folds considered here, a measurement of the tensor-to-scalar ratio consistent with zero then implies $\Delta\phi/m_\mathrm{Pl}\lesssim1.2$, meaning that a discrimination between large fields models and small fields models is possible for a wide range of values of $N_\mathrm{inf}$.

\begin{acknowledgments}
This research used resources of the National Energy Research Scientific Computing Center, which is supported by the Office of Science of the U.S. Department of Energy under Contract No. DE-AC02-05CH11231. Some of the results in this paper have been derived using {\sc s$^2$hat} \cite{s2hat,pures2hat,hupca_etal_2010,szydlarski_etal_2011}, {\sc healpix}~\cite{gorski_etal_2005} and {\sc class}\footnote{{\tt http://class-code.net}}~\cite{class} software packages.
\end{acknowledgments}

%%%%%%%%%%%%%%%%%%%%%%%%%%%%%%%%%%%%%%%%%%%%%%%%%

\end{document}